\documentclass[twocolumn]{aastex63}

\usepackage{natbib}
\usepackage{rotating}
\usepackage{graphicx}
\usepackage{diagbox}
\usepackage{multirow}
\usepackage{comment}
\usepackage{mathtools}
\usepackage{textcomp}
\usepackage[caption=false]{subfig}
\usepackage[toc,page]{appendix}
\usepackage{lineno}
\usepackage{enumitem}
\usepackage[english]{babel}  
\addto\extrasenglish{  
    
}
\usepackage{hyperref} 
\usepackage[hyphens]{url}
\usepackage{float}
\graphicspath{{./}{./figures/}}

\newcommand{\snrcom}[1]{\mbox{SNR$^{#1}$}}
\newcommand{\snrbmin}{\mbox{SNR$^b_{\mathrm{min}}$}}
\newcommand{\nviscom}[1]{\mbox{N$_{\mathrm{visits}}^{#1}$}}

\newcommand{\mfive}[1]{\mbox{m$_5^{#1,\mathrm{single}}$}}

\newcommand{\bg}{{$g$}}
\newcommand{\br}{{$r$}}
\newcommand{\bi}{{$i$}}
\newcommand{\bz}{{$z$}}
\newcommand{\by}{{$y$}}
\newcommand{\salt}{SALT2}

\newcommand{\strech}{$x_1$}
\newcommand{\snstrech}{\mbox{$x_1$}}
\newcommand{\col}{$c$}
\newcommand{\daymax}{$T_0$}
\newcommand{\sigc}{\mbox{$\sigma_c$}}
\newcommand{\sigstretch}{\mbox{$\sigma_{x_1}$}}

\newcommand{\zlim}{\mbox{$z_{\mathrm{lim}}$}}
\newcommand{\zlimfaint}{\mbox{$z_{\mathrm{lim,faint}}^{\mathrm{SN}}$}}
\newcommand{\cosmos}{{COSMOS}}
\newcommand{\elais}{{ELAIS-S1}}
\newcommand{\xmm}{{XMM-LSS}}
\newcommand{\cdfs}{{CDF-S}}
\newcommand{\adfa}{{ADF-A}}
\newcommand{\adfb}{{ADF-B}}
\newcommand{\adfs}{{Euclid/Roman}}
\newcommand{\euclid}{{Euclid}}
\newcommand{\romanspace}{{Roman Space Telescope}}

\newcommand{\sne}{{SNe~Ia}}
\newcommand{\snIa}{{SN~Ia}}
\newcommand{\sn}{{SNe}}
\newcommand{\degsq}{{deg$^2$}}
\newcommand{\nsn}{{N$_{\mathrm{SN}}^{z\leq z_{lim}}$}}
\newcommand{\nsntot}{{N$_{\mathrm{SN}}$}}
\newcommand{\nsnultra}{{N$_{\mathrm{SN}}^{\mathrm{Ultra}}$}}
\newcommand{\sigmansntot}{$\sigma_{\mathrm{N_{SN}}}$}
\newcommand{\nsncomp}{{N$_{\mathrm{SN}}^{z\leq z_{\mathrm{complete}}}$}}
\newcommand{\nsncompb}{{N$_{\mathrm{SN}}^{z\leq z_{\mathrm{complete}}}$}}

\newcommand{\zcomp}{\mbox{$z_{\mathrm{complete}}$}}
\newcommand{\zcompb}{\mbox{$z_{\mathrm{complete}}$}}

\newcommand{\sncolor}{\mbox{$c$}}

\newcommand{\per}{$\%$}
\newcommand{\seq}{$\sim$}
\newcommand{\nvisits}{$\mathrm{N_{visits}}$}
\newcommand{\nvisitsb}{\mbox{N$_{\mathrm{visits}}^b$}}

\newcommand{\nvisitsbmin}{\mbox{$\mathrm{N_{visits,min}^b}$}}

\newcommand{\nvisitsall}{\nviscom{g}, \nviscom{r}, \nviscom{i}, \nviscom{z}, \nviscom{y}}

\newcommand{\osfamily}[1]{{\it #1}}
\newcommand{\doffset}{tdo}
\newcommand{\fivesig}{\mbox{5-$\sigma$}}
\newcommand{\muth}{\mbox{$\mu_{th}(z,\Omega_m,w)$}}
\newcommand{\muthi}{\mbox{$\mu_{th}(z_i,\Omega_m,w)$}}
\newcommand{\sigint}{\mbox{$\sigma_{int}$}}
\newcommand{\sigintsq}{\mbox{$\sigma_{int}^2$}}
\newcommand{\sigsyst}{\mbox{$\sigma_{syst}$}}
\newcommand{\sigsystsq}{\mbox{$\sigma^2_{syst_i}$}}
\newcommand{\dew}{\mbox{$w$}}
\newcommand{\sigdew}{\mbox{$\sigma_{\dew}$}}
\newcommand{\dsigdew}{\mbox{$\Delta \sigma_\dew$}}
\newcommand{\deltasigdew}{\mbox{$\Delta \sigma_\dew$}}
\newcommand{\deltasigdewnsn}{\mbox{$\Delta \sigma_{\dew}^{\mathrm{N_{SN}}}$}}
\newcommand{\deltasigdewphotk}{\mbox{$\Delta \sigma_{\dew}^{\sigma_z^{k}}$}}
\newcommand{\deltasigdewphota}{\mbox{$\Delta \sigma_{\dew}^{\sigma_z^{0.002}}$}}
\newcommand{\deltasigdewphotb}{\mbox{$\Delta \sigma_{\dew}^{\sigma_z^{0.2}}$}}
\newcommand{\deltasigdewphotc}{\mbox{$\Delta \sigma_{\dew}^{\sigma_z^{0.1}}$}}
\newcommand{\mjdnight}{$\mathrm{MJD_{night}}$}
\newcommand{\duex}{DU}
\newcommand{\drall}{DR}
\newcommand{\edr}{IDR}
\newcommand{\ppm}{$\pm$}
\newcommand{\omgam}{$\Omega_m$}
\newcommand{\dus}[1]{DU$^{#1}$}
\newcommand{\drs}[2]{DR$_{#1}^{#2}$}
\newcommand{\edrs}[2]{IDR$_{#1}^{#2}$}
\newcommand{\pfs}{{PFS/Subaru}}
\newcommand{\tides}{{4MOST/TiDES}}
\newcommand{\nz}{{N($z$)}}
\newcommand{\photz}{{photo-$z$}}
\newcommand{\sigz}{{$\sigma_z$}}

\begin{document}

\title{Designing an Optimal LSST Deep Drilling Program for Cosmology with Type Ia Supernovae}


\author{Philippe Gris}
\affiliation{Laboratoire de Physique de Clermont, IN2P3/CNRS, F-63000 Clermont-Ferrand, France}
\author{Nicolas Regnault}
\affiliation{Laboratoire de Physique Nucléaire et des Hautes Energies, IN2P3/CNRS, France}
\author{Humna Awan}
\affiliation{Leinweber Center for Theoretical Physics, Department of Physics, University of Michigan, Ann Arbor, MI 48109, USA}
\author{Isobel Hook}
\affiliation{Physics Department, Lancaster University, Lancaster, United Kingdom LA1 4YB, United Kingdom}
\author{Saurabh~W.~Jha}
\affiliation{Department of Physics and Astronomy, Rutgers, the State University of New Jersey, Piscataway, NJ 08854, USA}
\author{Michelle Lochner}
\affiliation{Department of Physics and Astronomy, University of the Western Cape, Bellville, Cape Town, 7535, South Africa}
\affiliation{South African Radio Astronomy Observatory (SARAO), The Park, Park Road, Pinelands, Cape Town 7405, South Africa}
\author{Bruno Sanchez}
\affiliation{Department of Physics, Duke University, Durham, NC 27708, USA}
\author{Dan Scolnic}
\affiliation{Department of Physics, Duke University, Durham, NC 27708, USA}
\author{Mark Sullivan}
\affiliation{School of Physics and Astronomy, University of Southampton, Southampton, SO17 1BJ, UK}
\author{Peter Yoachim}
\affiliation{University of Washington, 4333 Brooklyn Ave NE, Seattle, WA 98105, USA}
\author{the LSST Dark Energy Science Collaboration}


\begin{abstract}

  The Vera C. Rubin Observatory's Legacy Survey of Space and Time is forecast to collect a large sample of Type Ia supernovae (\sne) that could be instrumental in unveiling 
  the nature of Dark Energy. The feat, however, requires measuring the two components of the Hubble diagram -~distance modulus and redshift~- with a high degree of accuracy. Distance is estimated from \sne~parameters extracted from light curve fits, where the average quality of light curves is primarily driven by survey parameters such as the cadence and the number of visits per band. An optimal observing strategy is thus critical for measuring cosmological parameters with high accuracy.
  We present in this paper a three-stage analysis aiming at quantifying the impact of the Deep Drilling (DD) strategy parameters (number of fields, cadence, number of seasons of observation, number of visits per band, time budget) on three critical aspects of the survey: the redshift completeness (originating from the Malmquist cosmological bias), the number of well-measured \sne, and the cosmological measurements. 
  Analyzing the current LSST survey simulations in a first stage, we demonstrate that the current DD survey plans are characterized by a low completeness (limited to $z~\sim$~0.55-0.65), and irregular and low cadences (few days) that dramatically decrease the size of the well-measured \sne~sample (by about 30\per). 
  We propose in a second stage a modus operandi that provides the number of visits (per band) required to reach higher redshifts. The results of this approach are used to design a set of optimized DD surveys for \sne~cosmology in a third stage. We show that most accurate cosmological measurements are achieved with Deep Rolling surveys characterized by a high cadence (one day), a rolling strategy (each field observed at least two seasons), and two sets of fields: ultra-deep ($z\gtrsim0.8$) and deep ($z\gtrsim0.6$) fields. We also demonstrate that a deterministic scheduler including a gap recovery mechanism is critical to achieve a high quality DD survey required for \sne~cosmology.

\end{abstract}




\section{Introduction}
\label{sec:intro}
Type Ia supernovae (\sne) are transient astronomical events resulting from a powerful and luminous explosion of a white dwarf. They display a characteristic brightness
evolution, with a luminosity peak about 15 days after explosion, and a slow decrease lasting up to few months. \sne~can be used as standardisable candles to determine cosmological distances. 
The Hubble diagram of \sne~is the most statistically efficient approach to constrain the dark energy equation of state (\citealt{Betoule_2014,Scolnic_2018}).
\par

The Vera C. Rubin Observatory’s Legacy Survey of Space and
Time (LSST \citealt{Ivezi__2019}) will discover few millions of supernovae during ten years of operations 
(\citealt{lsstsciencecollaboration2009lsst}). This number is quite impressive but in a sense misleading. If the survey is not optimized, a large fraction of these \sne~will be useless for cosmological measurements because of large luminosity distance errors. An optimized survey aims at {\it observing} a {\it large sample} (few thousands spanning a broad range of redshifts to be limited by systematic uncertainties) of {\it well-measured} \sne~with distance measurements accurate to better than 2-3$\%$. 

The ten-year LSST will image billions of objects in six bands. 80-90\per~of the observing time will be dedicated to Wide Fast Deep (WFD) primary survey, which will cover half of the sky ($\sim$ 18000 \degsq) at a universal\footnote{Fields are observed with a similar cadence and pattern.} cadence\footnote{The cadence is defined as the median inter-night gap in any filter.{\it High} cadences are characterized by {\it low} inter-night gaps.}. The remaining observing time will be shared among other programs (mini-surveys) including intensive scanning of a set of Deep Drilling (DD) fields. It is not clear yet what fraction of \sne~observed in the WFD (DD) survey will be confirmed from spectral features. But spectroscopically confirmed \sne~will certainly represent a small part of the \sne~sample. Accurate \sne~parameters will thus be estimated from well-measured light curves characterized by a sampling of few days and high Signal-to-Noise Ratio per band (\snrcom{b}).  
Obtaining these high quality light curves is therefore a key design point of the SN survey:  the average quality of the light curves depends primarily on the observing strategy.

In a recent paper (\citealt{lochner2021impact}), the LSST Dark Energy Science Collaboration (DESC) has presented an analysis of the WFD survey of observing strategies simulated by the LSST project\footnote{ \url{https://community.lsst.org/t/community-survey-strategy-highlights}.}. The conclusion is that an unprecedented number of high quality \sne~will be observed in the WFD survey (between 120k and 170k) up to redshifts $z~\sim$~0.3. The DD mini-survey of LSST is critical for observing a sample of high-redshift and well measured \sne~so as to achieve Stage IV dark energy goals (\citealt{albrecht2006report}). Optimizing the LSST DD survey so as to collect a large sample of well-measured \sne~up to high-redshift fulfilling this requirement while taking into account survey constraints (budget) is one of the main purpose of this paper. The work presented is a further step of a process started few years ago (\cite{scolnic_ddf_2018}, \cite{CadenceNote_2021}).

There are critical LSST survey parameters, such as scanning strategy, cadence, or filter allocation, that are not defined yet. Ongoing efforts are being made to define the requirements to accomplish the four primary science objectives of Rubin Observatory: dark energy and dark matter, inventory of the Solar System, transient optical sky exploration and mapping the Milky Way. As of 2020, the Survey Cadence Optimization Committee (SCOC \cite{Bianco_2021}) was charged to make specific recommendations for the survey parameter choices for LSST (initial survey strategy for ten years) based on input from the science community. The studies presented in this paper are part of the global effort aiming to define optimal strategy parameters to accomplish the scientific objectives of the Rubin Observatory.

\par
This paper deals with the interplay between the DD strategy and the \sne~sample collected by the survey. 
We perform a detailed study of the impact of the strategy parameters (number of fields to observe, number of seasons, season lengths, number of visits per night and per field) on the \sne~sample quality to assess whether observing supernovae up to $z~\simeq$~0.8-0.9 is achievable given design constraints, including in particular the number of visits alloted to DDFs. This article is subdivided into eight sections. The requirements for supernovae and the design constraints of the DD program are presented in \autoref{sec:reqsn} and \autoref{sec:design}. The metrics used to assess observing strategies are defined in \autoref{sec:metrics} and used in a detailed analysis of LSST simulations in \autoref{sec:analysis}. One of the conclusions of this analysis is that the samples collected with the proposed strategies could be too shallow and we propose in \autoref{sec:opti} a method aiming at increasing the depth of the DD survey. We use the results of this method to design optimized DD scenarios that would achieve the goal of observing high quality \sne~up to higher redshifts (\autoref{sec:scenario} to \autoref{sec:gaprecovery}).


\section{Requirements for supernovae} \label{sec:reqsn}
\sne~have been demonstrated as precise and reliable distance indicators in the last few decades.
Distances are derived from \sne~parameters that are infered from photometric light curves. The accuracy of the distance estimation reflects the precision of the photometric measurements. We discuss in the following (\autoref{sec:distmeas}) the light curve quality criteria that are required to obtain accurate distance measurements.

Supernovae surveys are magnitude limited and gather samples affected by a selection effect called the Malmquist bias (\citealt{Malmquist_1922}, \citealt{Teerikorpi_2015}, and references therein): brighter \sne~are preferentially discovered at the faint limits of the survey. This redshift-varying bias has an impact on the measurement of the cosmological parameters and is to be taken into account in the design of the survey (\autoref{sec:malmb}). 

\subsection{Distance measurement and well-measured supernovae} \label{sec:distmeas}
The diversity of \sne~light curves is usually parametrized by three parameters: an amplitude (brightness), a color and a light curve width (shape). The Tripp estimator (\citealt{Tripp_1999} and references therein) uses the B-band absolute magnitude, the (B-V) color, and the rate of decline during the first 15 days after maximum, $\Delta m_{15}$,  to standardize the \sne~brightness and estimate the distance. In the SALT2 light curve model (\citealt{Guy_2007,Guy_2010}), the distance modulus, $\mu$, is defined for each \snIa~by: 
\begin{equation} \label{eq:distmod}
    \mu = m_B + \alpha x_1 - \beta c -M
\end{equation}
where $m_B=-2.5~\mathrm{log_{10}}(x_0)+10.635$, where $x_0$ is the overall flux normalization, $x_1$ describes the width of the light curve, $c$ is a restframe color 
equal to $B$-$V$ at peak brightness. For each \snIa, $m_B,~x_1, ~c$ parameters are estimated from a fit of a \snIa~model to the measurements of a multicolor light curve. $\alpha$, $\beta$ and $M$ are global parameters estimated from the data. $M$ is the $B$ rest-frame magnitude. $\alpha$ and $\beta$ are global nuisance parameters quantifying the correlation between brightness with $x_1$ and $c$, respectively. The three parameters $\alpha,\beta,M$ are fitted along with cosmological parameters by minimizing the distance scatter. Accurate luminosity distances (i.e. accurate estimation of the \sne~standardisation parameters $m_B, x_1, c$) are thus critical to constrain cosmological parameters with better precision.
\begin{figure}[htbp]
\begin{center}
  \includegraphics[width=0.5\textwidth]{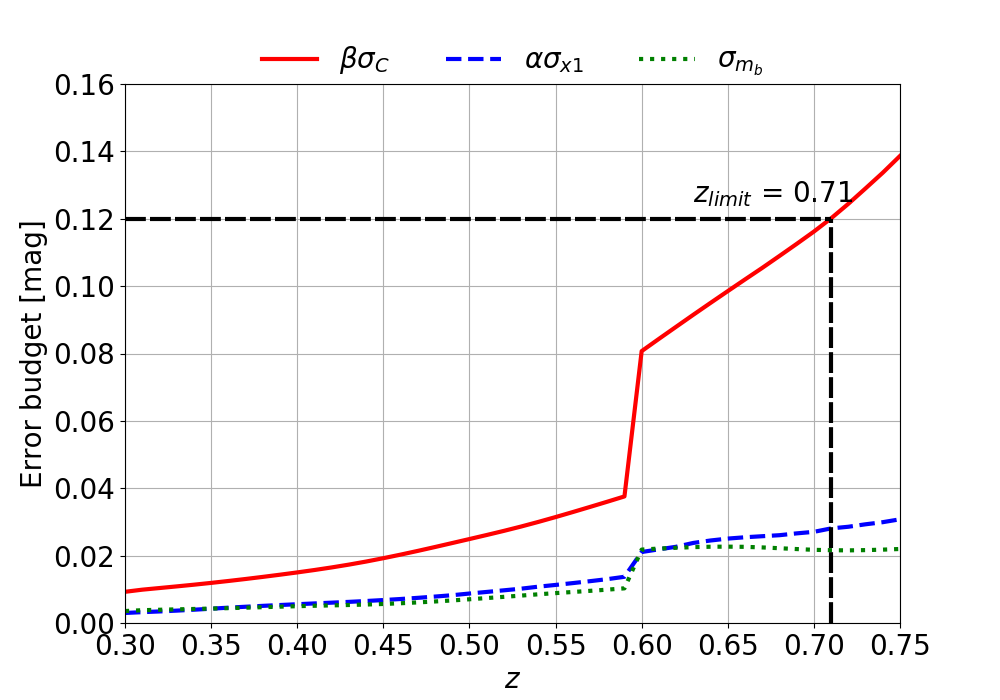}
 \caption{Contributions to the uncertainty of the distance modulus $\sigma_\mu$ as a function of the redshift. Three components are represented: the color component ($\beta \sigma_c$), the stretch component ($\alpha \sigma_{x_1}$) and the amplitude component ($\sigma_{m_b}$). The color component starts to contribute significantly to the distance uncertainty above $\beta\sigma_c~\sim$~0.12 (or equivalently $\sigma_c~\sim$ 0.04). This threshold corresponds to a redshift limit value (here 0.71) for the observation of well-measured \sne. These results were obtained from full simulation of \sne~light curves (regular cadence of one day). \sne~parameters were estimated from a SALT2 fit (see \autoref{sec:metrics} for more details).}\label{fig:errorbud_z}
\end{center}
\end{figure}

The relative contribution of the \snIa~parameter errors to the uncertainty on the distance modulus $\sigma_{\mu}$ is driven by the values of the nuisance parameters $\alpha$ and $\beta$. Recent measurements (\citealt{Scolnic_2018}, \citealt{Abbott_2019}) confirm that $\beta$ is larger than 3, that $\alpha$ is around 0.16 and that measured values of \strech~and \col~lie in limited ranges, $[-3.0,3.0]$ and $[-0.3,0.3]$, respectively. The consequence is that the color term $\beta \sigma_c$ is dominant in the $\sigma_{\mu}$ budget as illustrated in Fig. \ref{fig:errorbud_z}. The dispersion of Hubble residuals due to the intrinsic scatter of standardized \sne~brightness (\citealt{Brout_2021})
is of 0.12-0.14 mag (\citealt{Betoule_2014}, \citealt{Scolnic_2018}). The measurement uncertainties on the color above $\sim$~0.04 will thus make a significant contribution to the distance modulus errors. The requirement $\sigma_C~\lesssim$~0.04 is one of the main criteria (see \autoref{sec:metrics}) that designates a {\it well-measured} \snIa. It implicitly defines a redshift limit \zlim~ (Fig. \ref{fig:errorbud_z}) above which \sne~light curve measurements lead to inaccurate distance estimation and:
\begin{equation}
  \begin{aligned}
    \sigc \leq 0.04 & \Longrightarrow & \zlim
    \end{aligned}
 \label{eq:zlimsigmac}
\end{equation}

\begin{figure}[htbp]
\begin{center}
  \includegraphics[width=0.5\textwidth]{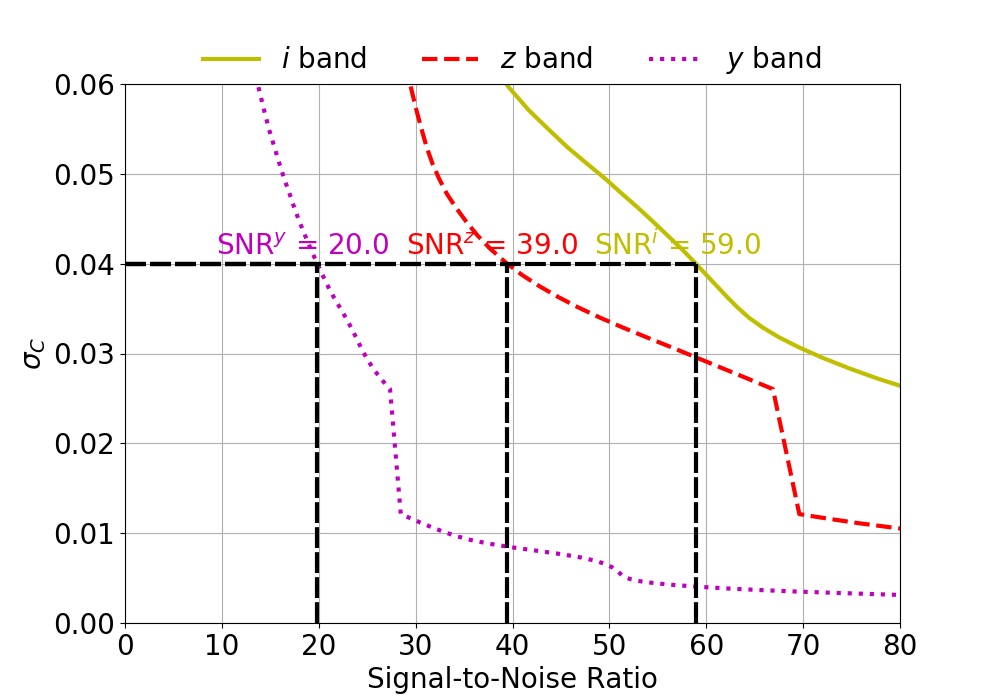}
 \caption{Color uncertainty as a function of the Signal-to-Noise Ratio per band. Requiring \sigc $\leq$ 0.04 is equivalent to applying the following selections: $SNR^i~\geq$~59 {\it and} $SNR^z~\geq$~39 {\it and} $SNR^y~\geq$~20. These results were obtained from full simulation of \sne~light curves for a medium \snIa~(regular cadence of one day). \sne~parameters were estimated from a SALT2 fit (see \autoref{sec:metrics} for more details)}\label{fig:sigmaC_z}
\end{center}
\end{figure}

The uncertainty on $m_b$, $x_1$, and $c$ is driven by the quality of the collected light curves which is defined by the sampling frequency of the measurements (cadence of observation) and by the light curve points uncertainties (observing conditions). \sigc~estimation is a function of the Signal-to-Noise Ratio (SNR) per band b, \snrcom{b},  defined by:
\begin{equation}
  \begin{aligned}
    \snrcom{b} &= \sqrt{\sum_{i=1}^{n^{b}}{\left(\frac{f_i^{b}}{\sigma_i^{b}}\right)^2}}
    \end{aligned}
  \label{eq:snrb}
\end{equation}
where $f^{b}$, and $\sigma^{b}$ are the fluxes and flux uncertainties. The summation runs over the number of light curve points. Requiring \sigc~$\leq$~0.04 is equivalent to requiring a minimal SNR per band (Fig. \ref{fig:sigmaC_z}) and the link between \zlim~and \snrcom{b}~may be written:
\begin{equation}
  \begin{aligned}
    \land (\snrcom{b} \geq \snrbmin) \Longrightarrow \sigc \leq 0.04 & \Longrightarrow \zlim
    \end{aligned}
 \label{eq:zlimsnr}
\end{equation}

where the logical symbol $\land$ means that the requirement $\snrcom{b} \geq \snrbmin$ is to be fulfilled for all the considered bands.
\subsection{Redshift completeness}
\label{sec:malmb}
As with all flux-limited surveys, 
a larger fraction of bright \sne~of the DD survey will systematically be observed at high-redshift.
\sne~observed at the fainter ends of the luminosity distribution are characterized by a mean intrinsic peak brightness higher than the mean of the whole sample. This bias increases with redshift and affects the distance estimation: the effective luminosity is biased towards brighter values. This leads to shorter distance measurements.

It is possible to estimate distance biases using simulation of the unobserved events (\citealt{Kessler_2013}, \citealt{Scolnic_2016}). Recent cosmological analysis (\citealt{Scolnic_2018}, \citealt{Riess_2019}) use the BEAM with Bias Corrections (BBC) framework (\citealt{Kessler_2017}) which includes corrections dependent on $\alpha$ and $\beta$ in the \sne~cosmology likelihood. With this method, the distance bias corrections have a clear impact on the cosmological measurements (the shift of the dark energy parameter \dew~decreases from 7$\%$ to 1$\%$) but the systematic uncertainty related to the selection bias still accounts for more than 20$\%$ of the total systematic error budget (\citealt{Scolnic_2018}). An incomplete high redshift sample is thus affected by a systematic uncertainty (due to selection bias) that could be dominant in high-redshift, magnitude-limited surveys like the LSST DD survey. More importantly, the Malmquist bias leads to a decrease of the number of \sne~for $z~\geq~\zcomp$: the fraction of higher redshift \sne~decreases with \zcomp. The redshift completeness value has thus an impact on \sne~cosmology since accurate cosmological measurements with a Hubble diagram heavily rely on the distribution of the number of well-measured \sne~as a function of the redshift, \nz. We will quantify this impact in \autoref{sec:scenario}. 

There are two ways to optimize the cosmological constraints from \sne~in a budget-limited survey. In a first scenario, the total number of well-sampled \sne~ can be maximized by observing all the DD fields for ten years. The second approach consists of maximizing the redshift completeness by observing the DD fields a limited number of years. We will study these two types of scenarios, (high \nsntot, low \zcomp) and (low \nsntot, high~\zcomp), in the following (\autoref{sec:scenario}). 



\section{Observing strategy constraints}
\label{sec:design}
The design parameters are the number of fields to be observed, the number of seasons of observation and the season length, the cadence of observation, the filter allocation, and the total observing time budget.

The Rubin Observatory defined in 2012\footnote{\url{http://ls.st/bki}.} four extragalactic Deep Drilling Fields:
\cosmos, \elais, \xmm, \cdfs~(Tab. \ref{tab:locddf}). More recently, the DESC collaboration has supported the LSST DDF coverage of the southern deep fields area (\citealt{CadenceNote_2021}) to ensure contemporaneous observations with \euclid~(\citealt{laureijs2011euclid,Amendola_2013}) and \romanspace~(\citealt{spergel2015widefield}), at the begining and at mid-term of the LSST survey, respectively.
\begin{table}[!htbp]
  \caption{Location of the DD fields considered in this study. AKARI Deep Fields (ADF-A and ADF-B) are examples of southern fields in the \adfs~area simulated in LSST observing strategy.}\label{tab:locddf}
  \begin{center}
    \begin{tabular}{c|c|c}
      \hline
      \hline
      Field & Central RA & Central Dec\\ 
      Name & (J2000)  & (J2000)\\
      \hline
       \elais & 00:37:48 & -44:01:30 \\
     \xmm & 02:22:18 &  -04:49:00 \\
    \cdfs & 03:31:55 & -28:07:00 \\
    \cosmos &10:00:26 & +02:14:01 \\
     \hline 
    \adfa & 04:51:00& -52:55:00 \\
    \adfb & 04:35:00 & -54:40:00 \\
      \hline
      \hline
      \end{tabular}
  \end{center}
\end{table}

\begin{figure}[!tbp]
\begin{center}
  \includegraphics[width=0.50\textwidth]{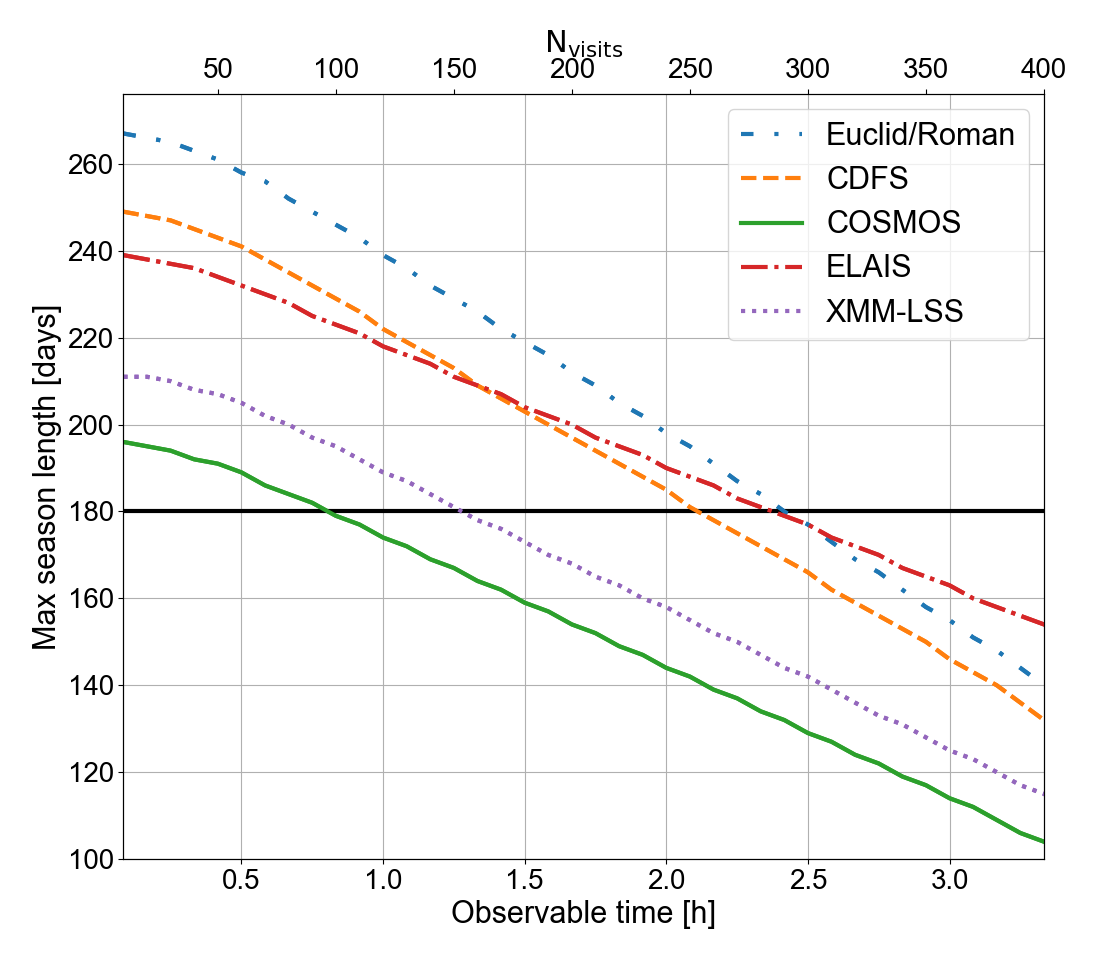}
 \caption{Maximal season length as a function of the nightly observable time (in hour) (lower $x$-axis) or the number of visits of 30$s$ (upper $x$-axis) per observing night. Fields are observable if the following requirements are met: 20\textdegree $\leq$ altitude $\leq$ 86.5\textdegree,  airmass $\leq$ 1.5. This plot was made using scripts and tools of the LSST scheduler. The black line corresponds to a season length of 180 days.}\label{fig:seasonlength_nvisits_new}
\end{center}
\end{figure}

The number of observed supernovae is proportional to the number of seasons of observation and to the season duration (\citealt{perrett}). 
The season length of a field is equal to the period of observability\footnote{An astronomical target is said to be observable if it is visible (i.e. for Rubin Observatory with altitude 20\textdegree~$\leq$ alt $\leq$ 86.5\textdegree and airmass $\leq$ 1.5) for a minimal amount of time.} which depends on its location w.r.t. the Vera C. Rubin Observatory. 
It is driven by the nightly observable time that can be converted to a number of visits of 30 s per observing night (\nvisits). The estimation of the season length as a function of the total number of visits for the fields defined in Tab. \ref{tab:locddf} (Fig. \ref{fig:seasonlength_nvisits_new}) suggests a decrease from 275-200 to 150-100 days when \nvisits~increases from 1 to 400. Season lengths of at least six months are required to collect at least 80\per~of \sne~of the northernmost fields. Maximizing season length is particularly important in the DDFs because of time dilation. High-$z$ \sne~light curves last longer than low-$z$ ones. \sne~collected at the beginning and at the end of the season are characterized by poorly reconstructed light curves leading to inaccurate distance measurements. Time dilation effects may be quantified as an effective season length decreasing with $z$. 

A regular cadence of observation ($\sim$~3 days max) is required to collect well-sampled light curves (LC). The number of large gaps ($>$~10 days) between visits degrades the 
measurements of luminosity distances, and potentially result in rejecting large sets of light curves of poor quality.

Measuring cosmological parameters with high accuracy requires observing \sne~over a wide redshift range $z\in[0.01,1.1]$. Five filters of the VRO have thus to be used: $g,r,i,z,y$ with a number of visits per band and per night depending on the redshift completeness of the survey (\autoref{sec:opti}).

\begin{figure}[htbp]
\begin{center}
  \includegraphics[width=0.5\textwidth]{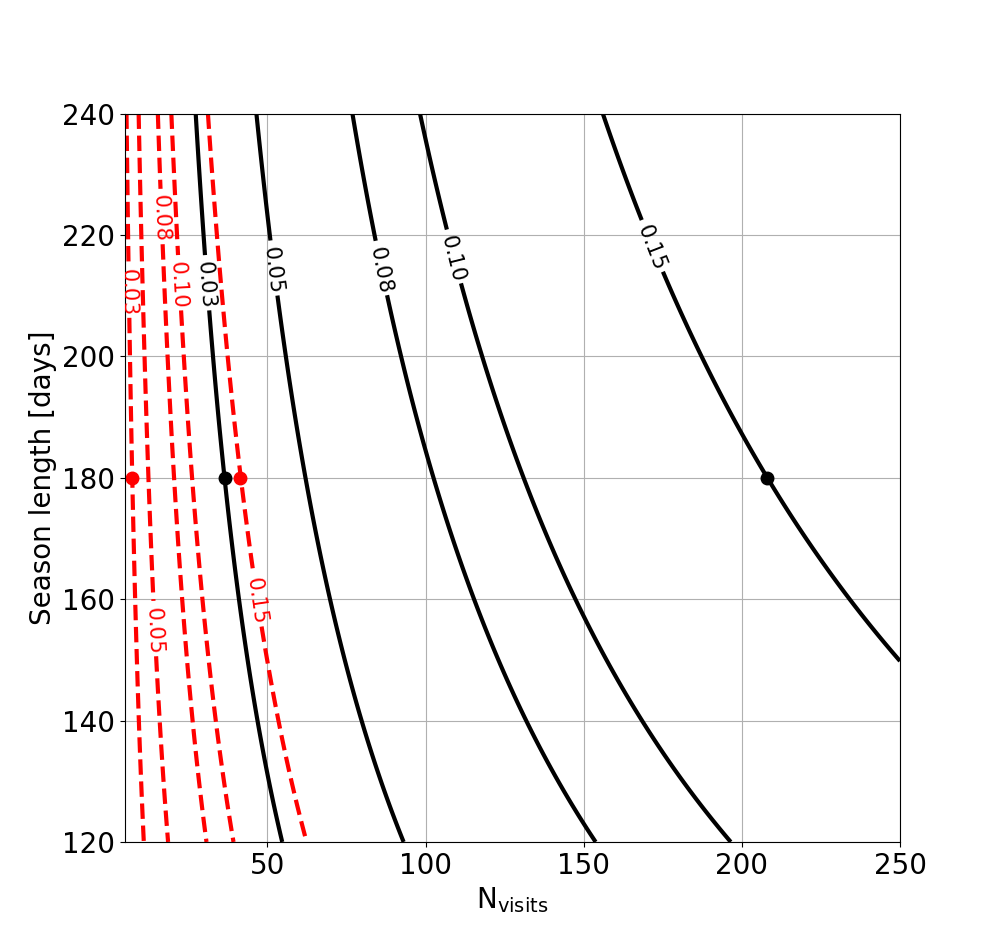}
 \caption{Budget contours in the plane (\nvisits, season length) for a configuration of 5 fields with two (black) or ten (red) observing seasons per field and a cadence of one day. The number of visits dramatically decreases with the number of seasons of observation when the time budget is limited. For a typical season length of 180 days and a budget extending from 3\per~(minimal) to 15\per~(maximal), the number of visits 
 ranges from 7 to 42 for 10 seasons (dashed red lines). Decreasing the number of seasons to 2 lead to an increase of the possible number of visits, from 36 to 208 (solid black lines).
 }\label{fig:bud_sl_nvisits}
\end{center}
\end{figure}

It is expected that 5-15$\%$ of the total number of LSST visits will be alloted to the DD program and shared among science topics interested by DD observations (such as AGN, supernovae, photo-$z$ training, ...).  The DD budget is defined as the fraction of observing time alloted to DDFs during the survey. For the sake of simplicity we will assume that the exposure time of observation does not change during the survey. In that case the budget is defined by:
\begin{equation}\label{eq:ddbudget}
\begin{aligned}
& \mathrm{DD_{budget} = N_{visits}^{DD}/(N_{visits}^{DD}+N_{visits}^{non-DD})}\\
\end{aligned}
 \end{equation}
where $N_{visits}^{DD}$ is the total number of visits (for the 10 years of Rubin Observatory operation) allocated to DDFs and is defined by:

\begin{equation}\label{eq:ddbudgetb}
\begin{aligned}
& \mathrm{N_{visits}^{DD}} = \sum_{i=1}^{\mathrm{N_{fields}}} \sum_{ j=1}^{\mathrm{N_{season}^i}} \mathrm{N_{visits,night}^{ij}\times seaslen^{ij}/cad^{ij} }\\
 \end{aligned}
 \end{equation}

where $\mathrm{N_{fields}}$ is the number of DD fields, $\mathrm{N_{season}}$ the number of season of observations per field, seaslen the season length (in days), cad the cadence of observation, and $\mathrm{N_{visits, night}^{ij}}$ the total number of visits per observing night, per field, and per season. The total number of visits corresponding to all fields but the DDFs, $N_{visits}^{non-DD}$, was estimated from a sample of LSST simulations and set to 2122176 visits (10 years of survey). The budget is fairly strongly dependent on the number of visits per observing night and on the season length (Fig. \ref{fig:bud_sl_nvisits}): the total number of visits is multiplied by 5 if the budget increases from 3\% to 15\%.

\section{Metrics to assess observing strategies}
\label{sec:metrics}
The metrics used to assess observing strategies are estimated from full simulation and fit of light curves. We have used SNCosmo\footnote{\url{https://sncosmo.readthedocs.io/en/latest/index.html}}(\citealt{Sncosmo_2016}), a Python library synthesising supernova spectra and photometry from SN models. It includes a lot of supernova models (SALT2, MLCS2k2, Hsiao, Nugent, PSNID, SNANA and Whalen models), as well as a variety of built-in bandpasses and magnitude systems. It includes functions for fitting and sampling SN model parameters given photometric light curve data. We have used the SALT2 model (\citealt{Guy_2007,Guy_2010}) where a \sne~is described by five parameters: $x_0$, \strech, \sncolor, $z$, and \daymax, the time of maximum luminosity. A flat-$\Lambda$CDM model was used to estimate cosmological distances, with $H_0$~=~ 70~km~s$^{-1}$, $\Omega_m$~=~0.3 and $\Omega_\Lambda$~=~0.7.
\par
In SALT2, model uncertainties of $g$ and $r$ bands (rest-frame UV) light curves fluxes are large (\citealt{Guy_2007}), and $g$ and $r$ observations with relative error model larger than 5$\%$ have not been considered in this study. This requirement implies that the list of filters useful to measure photometric light curves (observer-frame) is redshift-dependent (Tab. \ref{tab:zbands}).
\begin{table}[!htbp]
  \caption{List of filters useful to measure photometric light curves (observer-frame) as a function of the redshift.}\label{tab:zbands}
  \begin{center}
    \begin{tabular}{c|c|c|c|c}
      \hline
      \hline
      $z$ & [0.01,0.1] & [0.1,0.35] & [0.35,0.65] & [0.65,1.1] \\ 
      \hline
      bands &  \bg\br\bi & \bg\br\bi\bz& \br\bi\bz\by & \bi\bz\by\\
      \hline
      \hline
      \end{tabular}
  \end{center}
\end{table}
\par
Following the requirements for supernovae (\autoref{sec:reqsn}), we rely on two metrics to assess observing strategies: the redshift completeness \zcomp, and the number of well-measured \sne, \nsncomp. A well-measured \sne~is defined by the following tight selection criteria: 
\begin{itemize}
\item only light curve points with SNR~$\geq$~1 are considered; 
\item at least four (ten) epochs before (after) maximum luminosity are required, as well as at least one point with a phase\footnote{The phase of a LC point at time t is equal to $\frac{t-T_0}{1+z}$.} lower (higher) than -10 (20);
\item \sigc~$\leq$~0.04 is required to ensure accurate distance measurement. 
\end{itemize}
The redshift limit is defined as the maximum redshift of supernovae passing these selection criteria. 

The redshift of a complete sample, \zcompb, is estimated from the redshift limit distribution, \zlimfaint, of a simulated set of intrinsically faint supernovae (i.e. with ($x_1,~c$) = (-2.0, 0.2)) with \daymax~values spanning over the season duration of a group of observations. \zcompb~is defined as the 95th percentile of the \zlimfaint~cumulative distribution. 

The metrics are measured in HEALPix (\cite{Gorski_2005}) pixels of size 0.21~\degsq~over the region of the DDFs. For each pixel in the sky light curves are generated from observations of the simulated survey. Flux errors are estimated from 
the 5-$\sigma$ point source limiting magnitude (\fivesig~depth).
Light curves are fitted (using the \salt~model implemented in SNCosmo) to estimate \sne~parameters.

\section{Analysis of LSST simulations}
\label{sec:analysis}
LSST project has periodically released sets of simulations 
containing a large number of survey strategies. The simulations analyzed in this section were performed with the Feature-Based Scheduler (FBS \citealt{FBS_2019}) based on a modified Markov Decision Process that maximizes the scientific outcome of the Vera C. Rubin Observatory during its ten-year survey. It allows a flexible approach for scheduling. The sequential decisions of which filter and which pointing to select are estimated from features (weather conditions/image depth, slew time, footprint) to optimize observing strategy. The output of the simulations is composed by a set of observing parameters\footnote{The list of parameters is available at \\\url{https://github.com/lsst/sims\_featureScheduler}.} estimated at the center of the field of view of the telescope. These parameters serve as input for the generation of \sne~light curves (\autoref{sec:metrics}). 

The diversity of DD scenarios proposed in LSST simulations is rather limited and we have chosen to analyze a representative set of DD surveys on the basis of the following criteria: number of visits (and filter allocation) per observing night, cadence of observation, dithering, and budget. The list of LSST simulated observing strategies considered in this study is given in Tab. \ref{tab:os}. 

\begin{figure*}[htbp]
\begin{center}
  \includegraphics[width=0.99\textwidth]{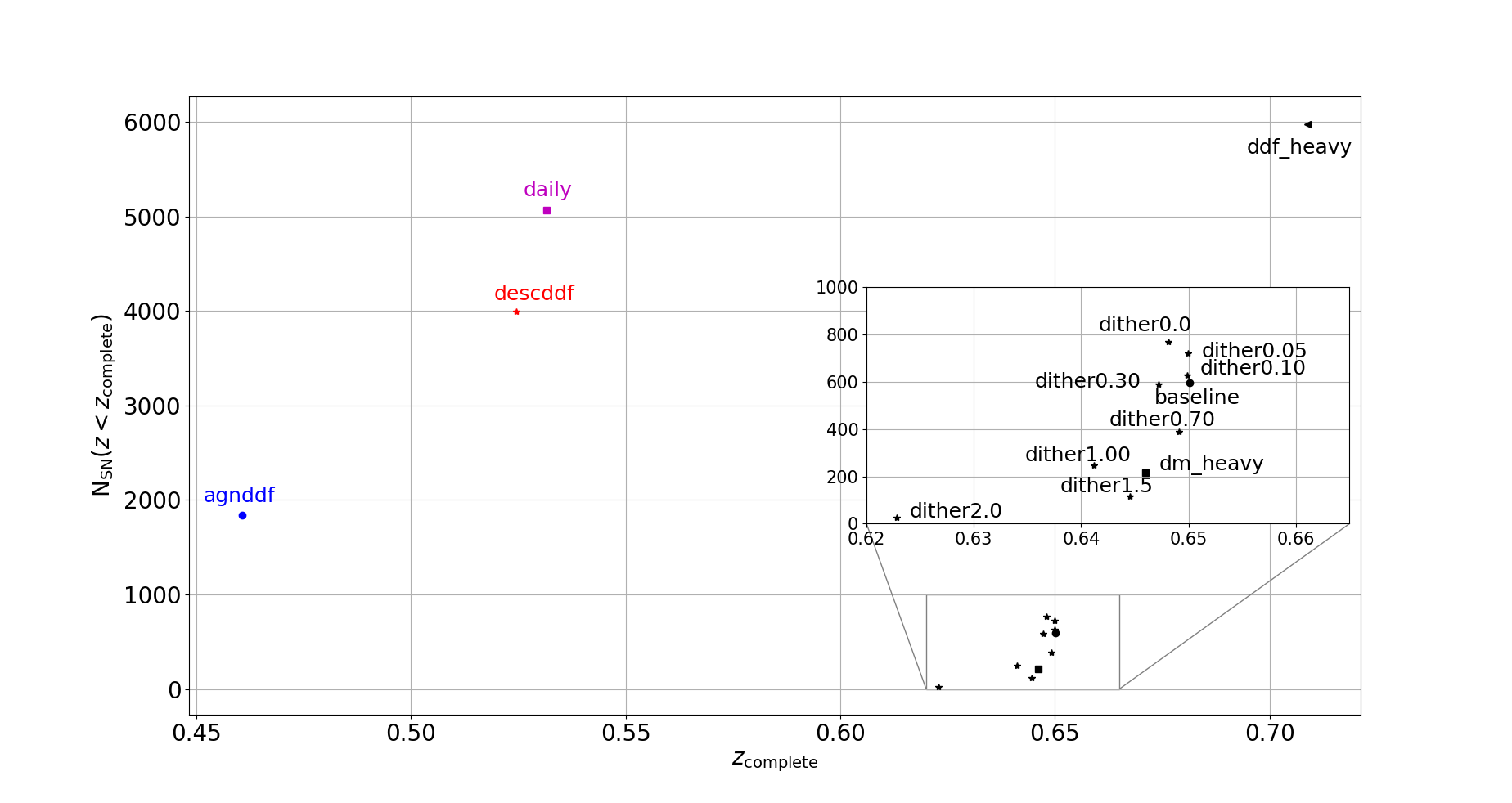}
 \caption{\nsncomp~vs \zcompb~for the LSST simulated observing strategies considered in this paper.}\label{fig:nsn_zlim_zoom}
\end{center}
\end{figure*}

\begin{table*}[htbp] 
\caption{Survey parameters for the list of observing strategies analyzed in this paper. For the cadence and season length, the numbers correspond to ADFS1/ADFS2/CDFS/COSMOS/ELAIS/XMM-LSS fields, respectivelly. The numbers following the filter allocation (\nvisits~column) are the minimum and maximum mean fraction of visits (per field over seasons) in the filter distribution. Only filter combinations with a contribution higher than 0.01 have been considered.}\label{tab:os} 
\begin{flushleft}
\hspace*{-2.5cm}
\scalebox{0.85}{
\begin{tabular}{c|c|c|c|c|c|c} 
  Observing & cadence & \nvisits & season length & area & DD budget & family\\ 
 Strategy & [days] & u/g/r/i/z/y & [days] & [deg2] &(\%)\\ 
\hline 
agnddf\_v1.5\_10yrs & 2.0/2.0/2.0/2.0/2.0/2.0 & -/1/1/3/5/4 [0.99-1.] & 164/165/235/189/171/177 & 112.9 & 3.4 & \osfamily{agn} \\ 
\hline 
baseline\_v1.5\_10yrs & 4.5/4.5/10.0/4.0/4.5/5.0 & -/10/20/20/26/20 [0.28-0.43] & 131/131/200/164/150/152 & 109.7 & 4.6 & \osfamily{baseline} \\
                                          &                                        & 8/10/20/20/-/20 [0.56-0.71] & & &  \\
\hline 
daily\_ddf\_v1.5\_10yrs & 2.0/2.0/2.0/2.0/2.0/2.0 & -/1/1/2/2/2 [0.60-0.61] & 161/161/236/188/171/178 & 113.5 & 5.5 & \osfamily{daily}\\
                                               &                                       & 1/1/1/2/-/2 [0.38-0.39] & & & \\
\hline 
ddf\_heavy\_v1.6\_10yrs & 2.0/2.0/2.0/2.0/2.0/2.0 & -/10/20/20/26/20 [0.26-0.39] & 116/116/201/167/152/150 & 110.6 & 13.4 & \osfamily{baseline}\\
                                                &                                       & 8/10/20/20/-/20 [0.60-0.72] & &  & \\
\hline 
&  & -/2/4/8/-/- [0.37-0.5] &  & & \\
descddf\_v1.5\_10yrs & 2.0/2.0/3.0/2.0/2.0/2.5 & -/-/-/-/25/4 [0.30-0.38] & 147/146/228/178/165/171 & 112.5 & 4.6 & \osfamily{desc}\\
 &  & -/-/-/-/-/4 [0.19-0.25] &  & & \\
\hline 
dm\_heavy\_v1.6\_10yrs & 7.5/6.0/14.0/8.5/8.0/7.0 & -/10/20/20/26/20 [0.31-0.45] & 119/119/195/142/139/138 & 188.6 & 4.6 & \osfamily{baseline}\\
                                                &                                        & 8/10/20/20/-/20 [0.54-0.68] &  &   &  \\
\hline 
ddf\_dither0.00\_v1.7\_10yrs & 4.0/4.0/6.0/2.0/3.0/3.0 & -/10/20/20/26/20 [0.17-0.43] & 121/123/204/165/153/159 & 69.2 & 4.6 & \osfamily{baseline}\\
                                                        &                                      & 16/10/20/20/-/20 [0.56-0.81] & 121/123/204/165/153/159 & 69.2 & 4.6 \\
\hline 
ddf\_dither0.05\_v1.7\_10yrs & 4.0/4.0/6.0/2.0/3.0/3.0 & -/10/20/20/26/20 [0.16-0.42] & 116/116/218/168/153/161 & 71.8 & 4.6 & \osfamily{baseline}\\
& & 16/10/20/20/-/20 [0.57-0.83] &&& \\ 
\hline
ddf\_dither0.10\_v1.7\_10yrs & 4.0/4.0/6.0/2.0/3.0/3.0 & -/10/20/20/26/20[0.19-0.43] & 120/120/220/165/150/165 & 74.7 & 4.6 & \osfamily{baseline}\\
& & 16/10/20/20/-/20 [0.57-0.81] &&& \\ 
\hline
ddf\_dither0.30\_v1.7\_10yrs & 4.0/4.0/6.5/3.0/3.0/3.0 & -/10/20/20/26/20 [0.21-0.45] & 118/118/201/167/146/146 & 83.5 & 4.6 & \osfamily{baseline}\\
& & 16/10/20/20/-/20 [0.54-0.78] &&& \\ 
\hline
ddf\_dither0.70\_v1.7\_10yrs & 4.5/4.5/9.0/4.0/4.0/4.25 & -/10/20/20/26/20 [0.19-0.43] & 123/137/201/163/146/146 & 104.5 & 4.6 & \osfamily{baseline}\\
& & 16/10/20/20/-/20 [0.57-0.79] &&& \\ 
\hline
ddf\_dither1.00\_v1.7\_10yrs & 4.0/4.0/14.0/5.5/5.0/5.0 & -/10/20/20/26/20 [0.23-0.43] & 113/118/198/153/143/143 & 124.4 & 4.6 & \osfamily{baseline}\\
& & 16/10/20/20/-/20 [0.56-0.77] &&& \\ 
\hline
ddf\_dither1.50\_v1.7\_10yrs & 4.5/4.5/16.5/8.5/6.75/6.0 & -/10/20/20/26/20 [0.20-0.42] & 121/121/196/145/135/139 & 159.3 & 4.6 & \osfamily{baseline}\\
& & 16/10/20/20/-/20 [0.57-0.79] &&& \\ 
\hline 
ddf\_dither2.00\_v1.7\_10yrs & 4.0/4.0/19.0/12.0/9.5/9.0 & -/10/20/20/26/20 [27-44] & 112/111/193/137/118/133 & 199.3 & 4.6 & \osfamily{baseline}\\
& & 16/10/20/20/-/20 [0.57-0.79] &&& \\ 
\end{tabular} 
}
\end{flushleft}
\end{table*}
Four sets of observing strategies can be defined from Tab. \ref{tab:os} according to the filter allocation per night, the parameter that has the most significant impact on \zcompb~value : \osfamily{baseline-like}~ (11 observing strategies), \osfamily{agn}, \osfamily{daily}, and \osfamily{desc} family. Estimation of the pair metric (\nsncomp, \zcompb) (defined in \autoref{sec:metrics}) for these families shows (Fig. \ref{fig:nsn_zlim_zoom}) that higher redshift limits are reached for the \osfamily{baseline-like} family. Most (10/11) of these observing strategies reach \zcompb~$\sim$~0.65. ddf\_heavy, the strategy with the largest DD budget, reaches \zcompb~$\sim$~0.72 and also collects the larger number of well-sampled \sne. \osfamily{daily} and \osfamily{desc} are characterized by a lower depth but by a significant number of well-measured \sne.\par

\begin{figure}[htbp]
  \includegraphics[width=0.52\textwidth]{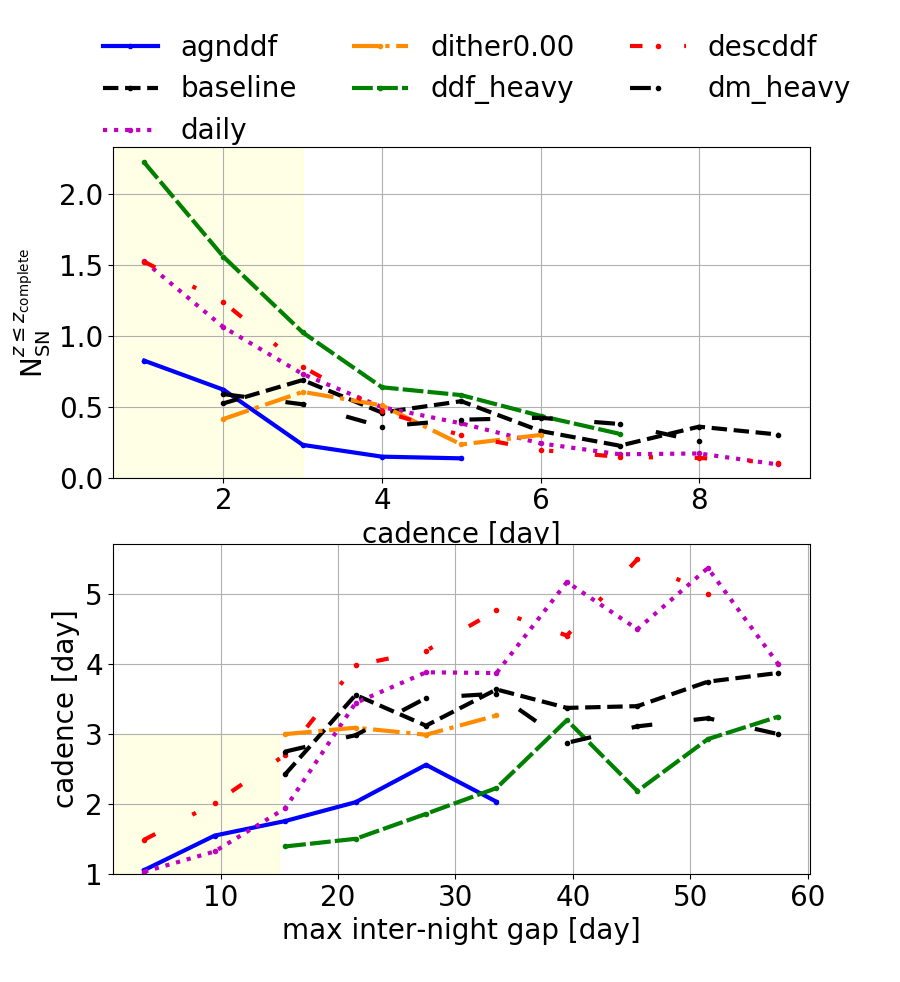}
 \caption{Median number of well-measured supernovae \nsncomp~as a function of the cadence of observation (top) and median cadence as a function of maximal inter-night gap (bottom) for a set of LSST simulated strategies studied in this paper. Yellow areas correspond to observing strategy parameters (cadence and max inter-nigh gaps) leading to a high quality \sne~sample.}\label{fig:nsn_cadence}
\end{figure}

The metric output (\nsncomp, \zcompb) is driven by the probability of a \sne~light curve to fulfill the requirements defined in \autoref{sec:metrics} (see \autoref{appendix:metric} for more details). This observing efficiency depends on the signal-to-noise ratio per band which is defined by the sampling frequency of the light curve points. The number of well-sampled \sne~is thus strongly dependent on the cadence of observation, as illustrated in Fig. \ref{fig:nsn_cadence} (top): as expected, higher cadences lead to higher (\nsncomp,~\zcompb).
   
Observing strategies studied in this paper are characterised by a wide range of cadences (Fig. \ref{fig:nsn_cadence}, top). Only two surveys, \osfamily{ddf\_heavy} and \osfamily{daily}, have more than 50\% of their observations with a 1-day cadence (Tab. \ref{tab:cadencesum}). \osfamily{baseline} and \osfamily{dither\_00}, two of the strategies with the lowest \nsncomp, present a large fraction of observations with a cadence of at least 3 days. These cadence distributions can be explained by some period with no observations. Two sources of gaps can be identified. One is the telescope downtime due to clouds and/or telescope maintenance. The other is the scanning strategy, when choices have to be made on which fields are to be observed on a given night. Inter-night gaps arising from telescope downtime lead to about 16-20$\%$ of nights without observation per season and are not expected to exceed few days (except for longer maintenance periods that could last up to 16 days see \autoref{sec:gaprecovery}). But the cadence may significantly increase for gaps higher than $\sim$ 10 days. Large gaps of few tens of days lead to a dramatic decrease of the cadence of observation, as illustrated in Fig. \ref{fig:nsn_cadence} (bottom).

\begin{table}[!htbp] 
\caption{Cadence distribution for a set of strategies studied in this paper. The cadence is estimated from nightly visits corresponding to all the filters $grizy$ for all the strategies but descddf which is characterized by observations related to $gri$ and $zy$ filters nightly interleaved (i.e. $gri$ visits one night and $zy$ visits the night after).}\label{tab:cadencesum} 
\begin{center} 
\begin{tabular}{c|c|c|c|c|c} 
\hline 
\hline 
\diagbox[innerwidth=2.cm,innerleftsep=-1.cm,height=3\line]{Strategy}{cadence} & 1-d & 2-d & 3-d & 4-d & $\geq$ 5-d\\ 
\hline 
agnddf & 37.6\% & 56.5\% & 5.0\% & 0.7\% & 0.1\% \\  
baseline & 0.0\% & 28.8\% & 44.7\% & 11.9\% & 14.6\% \\  
daily & 56.4\% & 21.2\% & 11.2\% & 5.6\% & 5.6\% \\  
dither0.00 & 0.0\% & 36.1\% & 43.7\% & 14.4\% & 5.8\% \\  
ddf\_heavy & 62.4\% & 22.8\% & 10.5\% & 3.2\% & 1.1\% \\  
descddf & 9.9\% & 58.9\% & 16.4\% & 7.0\% & 7.8\% \\  
dm\_heavy & 0.0\% & 31.0\% & 39.3\% & 18.4\% & 11.2\% \\  
\end{tabular} 
\end{center} 
\end{table}

\begin{figure}[htbp]
  \begin{center}
  \includegraphics[width=0.52\textwidth]{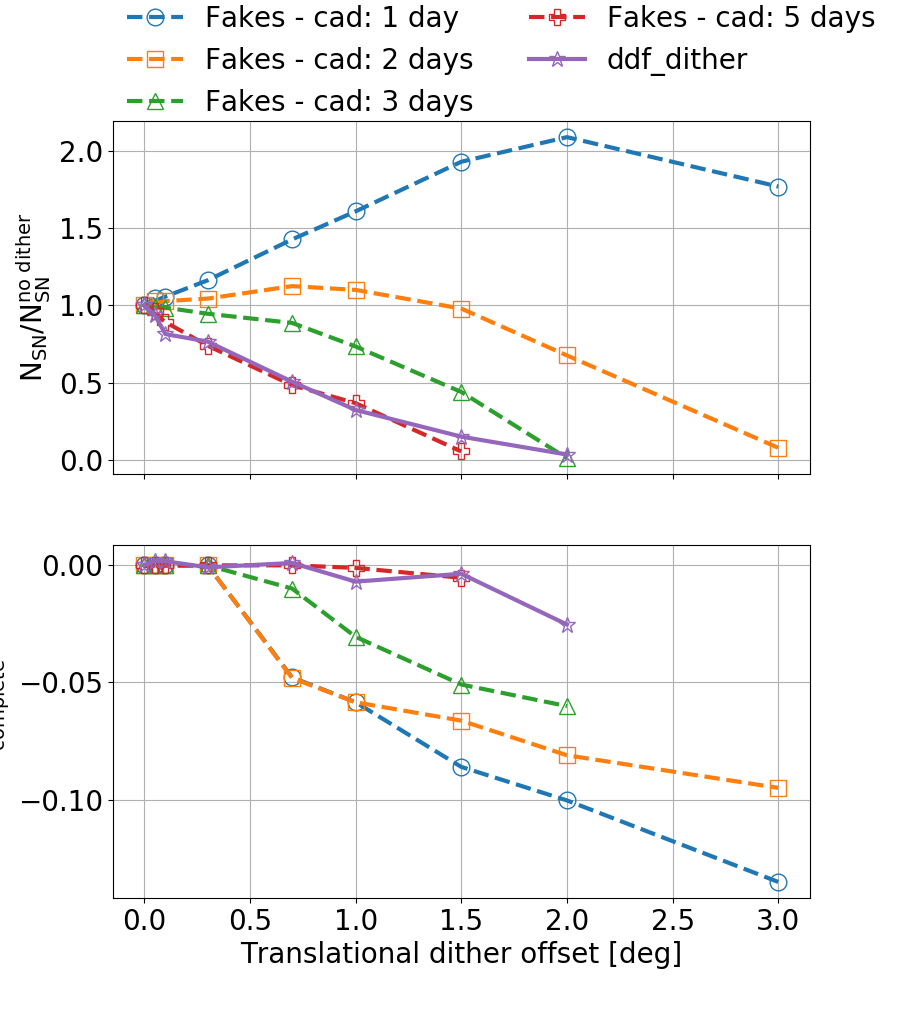}
 \caption{Ratio of the number of supernovae $N_{SN}/N_{SN}^{nodither}$ (top) and \zcompb~difference (bottom) as a function of translational dither offset. The simulations labelled as `Fakes' (dotted lines) correspond to regular cadences (1, 2, 3, 5 days) with median observing conditions (\fivesig~depth single exposure: 24.13/23.84/23.45/22.74/22.10 for $g/r/i/y/z$ bands, respectively.)}\label{fig:dither}
\end{center}
\end{figure}

The Rubin Observatory will provide a combination of large-scale dithers at each point of observation. Dithering patterns are composed of translational and rotational dithers \citep{Awan_2016}. The former corresponds to offsets of the telescope pointings, the latter to offsets of the camera rotational angles. We have studied the impact of the translational dithering on the metrics. It is expected to affect both the number of well-sampled supernovae and the redshift completeness for each of the pixels of the  field. With no dithering, \nsncomp~and \zcompb~distributions are uniform across the whole field area. The translational dithering has an impact on edge pixels which are thus characterized by a lower cadence w.r.t. central pixels.
A decrease of both \nsncomp~and \zcompb~(per pixel) is then observed for edge pixels. 
\zcompb~tends to decrease with an increase of the translational dither offset (\doffset), with a greater effect for high cadences. The number of supernovae is the result of the trade-off between two effects (Fig. \ref{fig:dither}): an increase of the survey area (which increases with \doffset) and a decrease of the cadence ( which decreases with \doffset). The increase of the survey area lead to an increase of the number of supernovae for high cadences and low \doffset~values.

In summary, the LSST simulated strategies lead to the observation of a sample of well-measured \sne~with a rather low \zcompb, the redshift limit of the complete cosmology-grade \sne~sample. It will be shown in the following (\autoref{sec:scenario}) that reaching \zcompb~$\sim$~0.6 is not sufficient to achieve a measurement of \dew~with a high degree of precision.
\zcomp~is mainly driven by \snrcom{b}~values (Eq. \ref{eq:snrb}) which depends on the number of visits in the corresponding band \nvisitsb~(\autoref{sec:opti}). 
The number of visits per band has thus to be increased to reach higher \zcomp. We propose in \autoref{sec:opti} a method for assessing the relationship between \nvisitsb~and \zcomp. The second conclusion of these studies is that the cadence is a key parameter to collect a large sample of well-measured \sne. High cadences are favored to maximize \snrcom{b}. Large inter-night gaps are harmful as they lead to a decrease of the number of well-measured supernovae. It is critical to reduce the inter-night gaps originating from the survey strategy so as to maximize the size and depth of the well-measured \sne~sample. Finally, larger translational dithers reduce the DDF area with high cadence and lead to a dramatic decrease of the number of well-sampled \sne~for low cadence~($\gtrsim$~3~days) strategies.

\section{Optimization of the number of visits}
\label{sec:opti}

The analysis of LSST simulations has shown (see \autoref{sec:analysis}) that it seems difficult to collect complete samples of \sne~with redshift higher than \zcompb~\seq~0.55-0.65. The proposed cadences of observation, filter allocation and season lengths do not allow to reach higher redshifts for a DD budget of \seq~5\per. According to Eq. \ref{eq:zlimsnr}, reaching higher \zcomp~requires increasing \snrcom{b}. 

The signal-to-noise ratio per band is the complex result of the combination of the \snIa~flux distribution ($z$-dependent), the number of visits per band, the cadence of observation, and observing conditions (5-$\sigma$ depth). It is thus not possible to estimate the 
observing strategy parameters required to reach higher redshifts from the results of \autoref{sec:analysis} (by a simple rescaling for instance). This is why we present in this section a study to assess the relationship between the redshift completeness and the number of visits per band and per observing night (for a defined cadence). The optimized number of visits per band required to reach higher redshifts estimated with this approach is a key parameter to build DD scenarios consistent with the list of constraints presented in \autoref{sec:reqsn} and in \autoref{sec:design}.
\par
As described in Eq. \ref{eq:ddbudget} the DD budget depends primarily on 5 parameters: the number of fields to observe, the season length (per field and per season), the number of seasons of observation (per field), the cadence of observation (per field and per season), and the number of visits \nvisitsb~per filter and per observing night. \nvisitsb~is related to \snrcom{b}~through the flux measurement uncertainties  $\sigma_i^b$. In the background-dominated regime one has $\sigma_i^b \simeq \sigma_5^b$ where $\sigma_5^b$ is equal by definition to
\begin{equation}\label{eq:opt2}
  \begin{aligned}
    \sigma_5^b &=  \frac{f_5^b}{5}
    \end{aligned}
\end{equation}
where $f_ 5^b$ is the \fivesig~flux related to the \fivesig~depth magnitude $m_5^b$ through:
\begin{equation}\label{eq:opt3}
  \begin{aligned}
    m_5^b &= -2.5 \log f_5^b+zp^b
    \end{aligned}
\end{equation}
where $zp^b$ is the zero point of the considered filter.  $m_5^b$ is related to \nvisitsb through:
\begin{equation}\label{eq:opt4}
  \begin{aligned}
    m_5^b - m_5^{b, \mathrm{single}} & \approx  1.25 \log(N_{visits}^b)
    \end{aligned}
\end{equation}
where $m_5^{b, \mathrm{single}}$ is the \fivesig~depth corresponding to a single visit, a parameter depending on observing conditions. These equations \eqref{eq:opt2}-\eqref{eq:opt4} describe the relationship between \snrcom{b}~ and \nvisitsb. The requirement $\snrcom{b}~\geq~\snrbmin$ is equivalent to $\nvisitsb~\geq~\nvisitsbmin$ and Eq. \eqref{eq:zlimsnr} may be written:
\begin{equation}
  \begin{aligned}
    \land(\nvisitsb~\geq~\nvisitsbmin) \Longrightarrow \sigc \leq 0.04 \Longrightarrow \zlim
    \end{aligned}
 \label{eq:zlimnvisits}
\end{equation}
\\
where the logical symbol $\land$ means that the requirement $\nvisitsb~\geq~\nvisitsbmin$ is to be fulfilled for all the considered bands. The relations \eqref{eq:zlimsnr} and \eqref{eq:zlimnvisits} are not univocal. Many \snrcom{b}~(\nvisitsb) combinations lead in fact to the same result and constraints have to be applied to choose optimal configurations. 

\begin{figure}[!tbp]
    \includegraphics[width=0.5\textwidth]{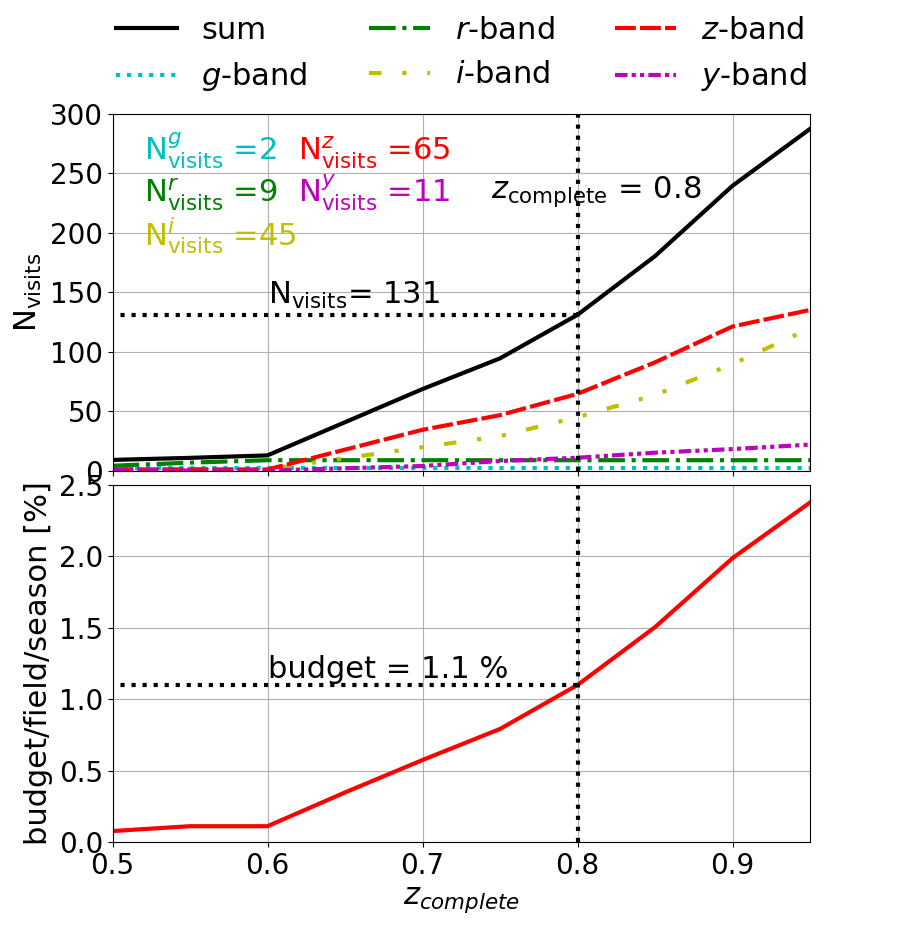}
  \caption{Top: number of visits per observing night as a function of the redshift completeness. 131 visits {\it with} the following filter allocation (\nvisitsall)=(2, 9, 45, 64, 11) are required per observing night to reach \zcomp\seq 0.8 for a cadence of one day. Bottom: budget per field and per season of observation as a function of \zcomp. A 1-day cadence and a season length of 180 days have been assumed. Observing a field every night for 180 days with a redshift completeness of 0.80 corresponds to a budget of 1.1\per.}\label{fig:nvisits_zlim_cadence}
  
\end{figure}

We have used the following method to estimate N$_{\mathrm{visits}}^b(z)$. A systematic scan of the SNR parameter space (\snrcom{g}, \snrcom{r}, \snrcom{i}, \snrcom{z}, \snrcom{y}) is performed. Median observing conditions estimated from DD simulations are used, namely \mfive{g} = 24.48, \mfive{r} = 23.60, \mfive{i} = 24.03, \mfive{z} = 22.97, \mfive{y} = 22.14. For each SNR combination and for a set of cadences (1 day to 4 days), light curves of an intrinsically faint \sne~(i.e. with \snstrech~=~-2.0, \sncolor~=~0.2) in the redshift range [0.01,1.0] are simulated using templates and \sne~parameter errors (\sigc, \sigstretch) are estimated using the Fisher Matrix formalism. This approach considerably reduces the processing time (compared to full simulation+fit) while ensuring the highest degree of accuracy of the LC points (fluxes and flux errors) and of the supernovae parameter errors. The light curves fulfilling the requirements defined in \autoref{sec:metrics} are used to define the SNR parameter space, or equivalently the \nvisits~ parameter space (\nvisitsall)~according to Eq. \ref{eq:zlimnvisits}, corresponding to well-measured \sne. Optimal combination are selected by minimizing the total number of visits per observing night and by requiring a maximum number of \by-band visits. This selection aims at reducing the (potentially severe) systematic effects affecting the \by-band measurements \citep{High_2010}. The result is displayed in Fig. \ref{fig:nvisits_zlim_cadence} (top) for a 1 day cadence. The number of visits strongly increases with the redshift completeness for \zcomp~$\gtrsim~0.7$ where only three bands \bi\bz\by~can be used to construct \sne~light curves. About 130 visits (1 hour and 5 minutes of observation) are required to reach \zcomp~$\sim$~0.8 for a one day cadence. Since the number of visits required to reach a given \zcomp~value increases linearly (as a first approximation) with the cadence, this corresponds to $\sim$~3.25~hours of exposure time per night for a 3-day cadence.

It is known that the restframe UV region is subject to large fluctuations between \sne~in the \salt~light curve model (see \autoref{sec:metrics}). One of the consequences is that only three (two) bands, $i,z,y$ ($z,y$), may be used to reconstruct light curves for redshifts higher than $\sim$~0.7 ($\sim$ 1.1). The \sne~parameter errors depend on the SNR values, but also on the shape of the light curves per band. The contribution of \snrcom{y}~ to the \sne~parameter errors tends to increase with \zcomp~if the total number of visits remains relatively constant. A high number of \nviscom{y}~for low \zcomp ($\sim$~0.7) leads to a net loss of well-sampled \sne~because of bad color measurements. The optimal \nviscom{y}~as a function of \zcomp~has been estimated by computing the redshift limit of a medium \sne~in configurations with 5$~\leq~$\nviscom{y}$~\leq~$80. Requiring a \zlim~variation lower than 0.01 leads to the results of Tab. \ref{tab:Ny_optim}. 

\begin{table}[!htbp]
  \caption{A set of optimal numbers of $y$-band visits as a function of \zcomp.}\label{tab:Ny_optim}
  \begin{center}
    \begin{tabular}{c|c|c|c|c|c|c}
      \hline
      \hline
      \zcomp & 0.65 & 0.70 & 0.75 & 0.80 & 0.85 & 0.90 \\
      \hline
      \nviscom{y} & 3 & 7 & 16 & 21 & 30 & 38 \\
      \hline
    \end{tabular}
  \end{center}
\end{table}

It is possible to estimate the budget per field and per season of observation as a function of \zcomp~by using Eq. \ref{eq:ddbudget} and the results of Fig. \ref{fig:nvisits_zlim_cadence} (top). As expected (Fig. \ref{fig:nvisits_zlim_cadence}, bottom) a significant increase is observed for higher redshifts and the observation of a field for 180 days with \zcomp$~\sim~$0.9 requires a budget of 2\per.

The optimized number of visits required to reach higher redshift completeness is the last piece of the puzzle to be included in the budget estimator (Eq. \ref{eq:ddbudget}). We have now the tools to design realistic and optimal DD scenarios.

\section{Optimal LSST DD surveys for cosmology with \sne}
\label{sec:scenario}
In the following, we examine three key points of the surveys (redshift completeness, cadence of observation, and cosmological measurements) before presenting a set of optimized scenario.

\paragraph{Redshift completeness}
Spectroscopic datasets from cosmological endeavors overlapping with LSST in area and timing are essential for \sne~cosmology. They provide enormous added benefits through (a) the follow-up of a subset of the full sample of well-measured supernovae (to improve the models used to build training sample for photometric classification), and (b) the measurement of host-galaxy redshifts with high accuracy. Three of the spectroscopic resources contemporaneous with LSST, Dark Energy Spectroscopic Instrument (DESI \citealt{desicollaboration2016desi}), Primary Focus Spectrograph (\pfs~\citealt{Tamura_2016}) and 4MOST (\citealt{4MOST}), will provide vital live spectra and host spectroscopic redshifts (\citealt{mandelbaum2019widefield}).

DESI is a ground-based dark energy experiment used to conduct a five-year survey that will cover 14,000 \degsq. More than 30 million galaxy and quasar redshifts will be measured to study baryon acoustic oscillation and the growth of structure through redshift-space distorsions. The DESI survey will overlap with at least 4000 \degsq~of the LSST footprint in the northern hemisphere.

The \pfs~spectroscopic follow-up survey is designed to observe two of LSST DDFs accessible from the Subaru telescope: \cosmos~and \xmm. About 2000 spectra of live supernovae and 20,000 host galaxy redshifts up to $z\sim0.8$ will be collected after 10 years. 

The 4MOST Time-Domain Extragalactic Survey (TiDES \citealt{TiDES}) is dedicated to the spectroscopic follow-up of extragalactic optical transients and variable sources selected from e.g. LSST. The goal is to collect spectra for up to 30,000 live transients to $z\sim0.5$ and to measure up to 50,000 host galaxy redshifts up to $z\sim1$. This corresponds to both the DD and the WFD fields. 

Two sets of LSST fields may then be defined to fully benefit from the synergy with DESI, \pfs~and 4MOST. DESI and \pfs~will provide live-spectra and spectroscopic redshifts for the northernmost fields, \cosmos~and \xmm, over a broad range in redshift. Southern fields, \adfs, \cdfs, \elais, will take advantage from \tides~ measurements.

\paragraph{Cadence of observation}

Few arguments point in favour of high cadences: the total number of visits per night, the season length, the translational dithering, and inter-night gaps.

More than 240 visits are required to reach \zcomp~$\geq$~0.9 for a 1-day cadence (Fig. \ref{fig:nvisits_zlim_cadence}, top). The same depth is obtained for a 3-day cadence with more than 720 visits, that is about 6 hours of observation. Reaching higher \zcomp~with low cadence observing strategies is thus not realistic: it would potentially jeopardize the uniformity of the WFD survey. 

Because of their northernmost positions, \cosmos~and \xmm~are characterized by shorter season lengths w.r.t. other fields (for the same observing time per night) (Fig. \ref{fig:seasonlength_nvisits_new}). Requiring more than 480 visits per night (to reach \zcomp~$\geq$~0.9 with a 2d cadence) would dramatically degrade the season lengths (to less than 90 days for these two fields) and would drastically reduce the size of the well-measured supernovae sample (by more than 50 $\%$).

As shown on Fig. \ref{fig:dither} the translational dithering has a limited impact on the number of well-measured \sne~and on \zcomp~up to tdo$\sim$1 degree for high cadences. The number of well-measured \sne~is falling rapidly with tdo for cadences lower than 3 days.

One of the main conclusions of the analysis of the proposed LSST simulations (\autoref{sec:analysis}) is that large inter-night gaps have harmful effects on the sampling and on the quality of \sne~light curves. They can be reduced to a minimum (i.e. to unavoidable gaps related to telescope maintenance or bad weather conditions) by either observing DDFs at high cadences or including a mechanism in the scheduler that would ensure to keep a high observing rate of \sne~(see \autoref{sec:gaprecovery} for suggestions).

\paragraph{Cosmological metric}
The most accurate way to measure cosmological parameters from a sample of well-measured \sne~is to perform a maximum likelihood analysis by minimizing:
\begin{equation}\label{cosmofit}
    -ln \mathcal{L} = ( \mu-\muth)^\intercal C (\mu-\muth)
\end{equation}
where $C$ is the covariance matrix and $\mu$ is the distance modulus (Eq. \ref{eq:distmod}). \muth~is defined by:
\begin{equation}
    \muth = 5 \log_{10}[d_L(Mpc)]+25
\end{equation}
In a flat universe, the luminosity distance $d_L$ ~is defined by:

\begin{equation}
    d_L(z,\Omega_m,w)= {\frac{c(z+1)}{H_0}}\int_0^z{\frac{dz^{'}}{\sqrt{(1-\Omega_m)+\Omega_m(1+z^{'})^{3\dew}}}}
\end{equation}
where $\Omega_m$ the dark matter density parameter, and \dew~is the parameter of the dark energy equation of state. Five parameters are to be estimated: the cosmological parameters ($\Omega_m,\dew$) and the nuisance parameters ($M,\alpha,\beta$) (Eq. \ref{eq:distmod}).

The goal of this section is to study a large set of surveys by varying the number of fields to be observed, the redshift completeness (i.e. having samples with redshift completeness field-dependent) and the number of seasons. Using the above-mentioned method for each scenario would require to produce a lot of (time-consuming) simulations and fits of \sne~light curves to estimate cosmological parameters. We have thus chosen to work with distance moduli of supernovae simulated using:
\begin{equation}
    \mu_i(z_i) \sim \mathcal{N}(\muthi, \sigma^2 = \sigma_{\mu_i}^2+\sigintsq+\sigsystsq)
\end{equation}
where $\sigint$ is the intrinsic dispersion of supernovae ($\sigint \sim$ 0.12) and \sigsyst~accounts for systematic uncertainties. $\sigma_\mu(z)$ is the distance modulus error for each \sne~of redshift $z_i$. It has been estimated from a complete simulation of DD surveys with varying \zcomp~using the method developed in \autoref{sec:metrics}. 

The cosmological parameters are estimated by minimizing:  
\begin{equation} \label{eq:chi_square}
-ln\mathcal{L}= \sum_{i=1}^{N_{SN}} \frac{(\mu_i-\muthi)^2}{\sigma_{\mu_i}^2+\sigintsq+\sigsystsq}
\end{equation}
where $N_{SN}$ is the number of well-measured \sne~used to perform the fit. The realistic simulations used to estimate $\sigma_\mu$ and $N_{SN}$ take into account redshift bias (see \autoref{appendix:realsimu} for more details).
For each DD scenario considered in the following, a sample of about 100,000 low-$z$ \sne~(10000 per year) has been added up to $\zcomp\sim$0.2 \citep{lochner2021impact}.

The following sections describe a set of optimized scenarios based on two different approaches. One is optimizing the number of well-sampled \sne~collected by the survey. The other aims at probing high redshift completeness domains. A survey is characterized by three parameters, the redshift completeness, the number of DDFs, and the cadence of observation. Three metrics are presented to assess the proposed scenario: the DD budget, the cosmological metric (we have chosen the error on the \dew~parameter, 
$\sigma_{\dew}$), and the total number of well-sampled \sne.

\subsection{Deep Universal Surveys} \label{sec:deep_universal_survey}
\begin{figure*}[!tbp]
  \begin{minipage}[b]{0.9\textwidth}
    \includegraphics[width=\textwidth]{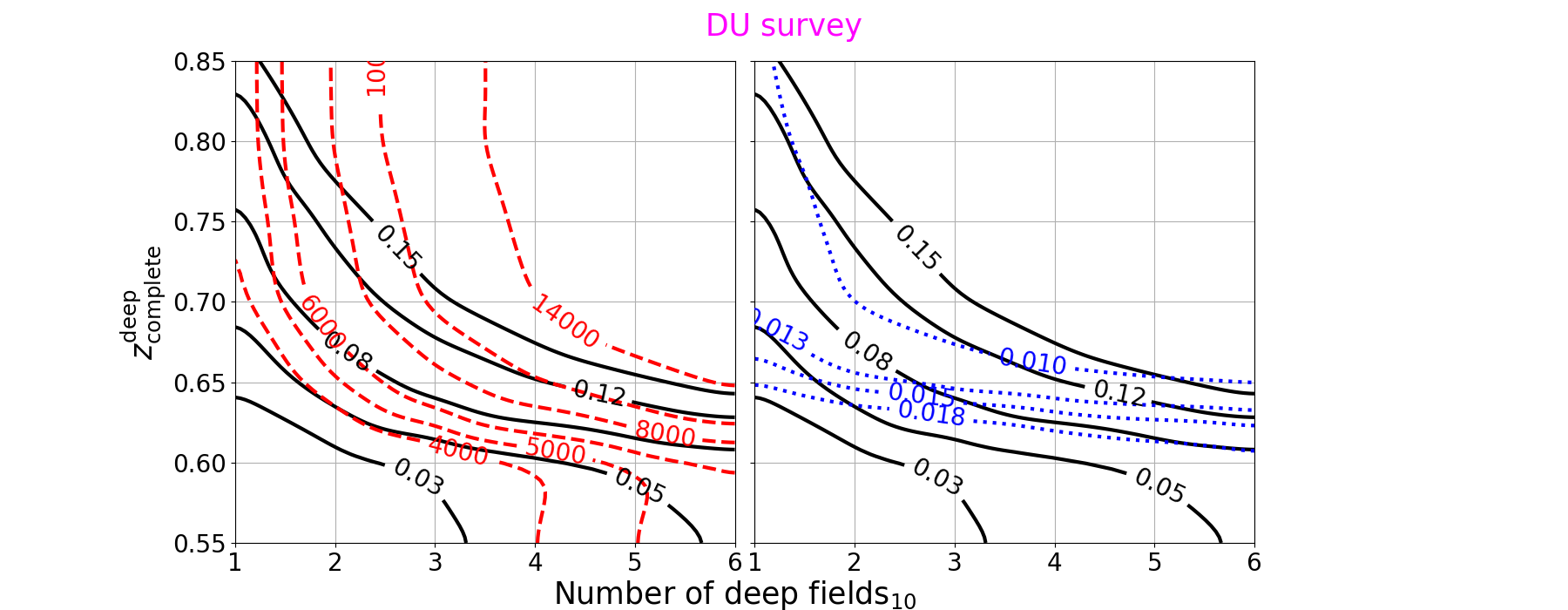}
  \end{minipage}
 \hfill
    \begin{minipage}[b]{0.9\textwidth}
    \includegraphics[width=\textwidth]{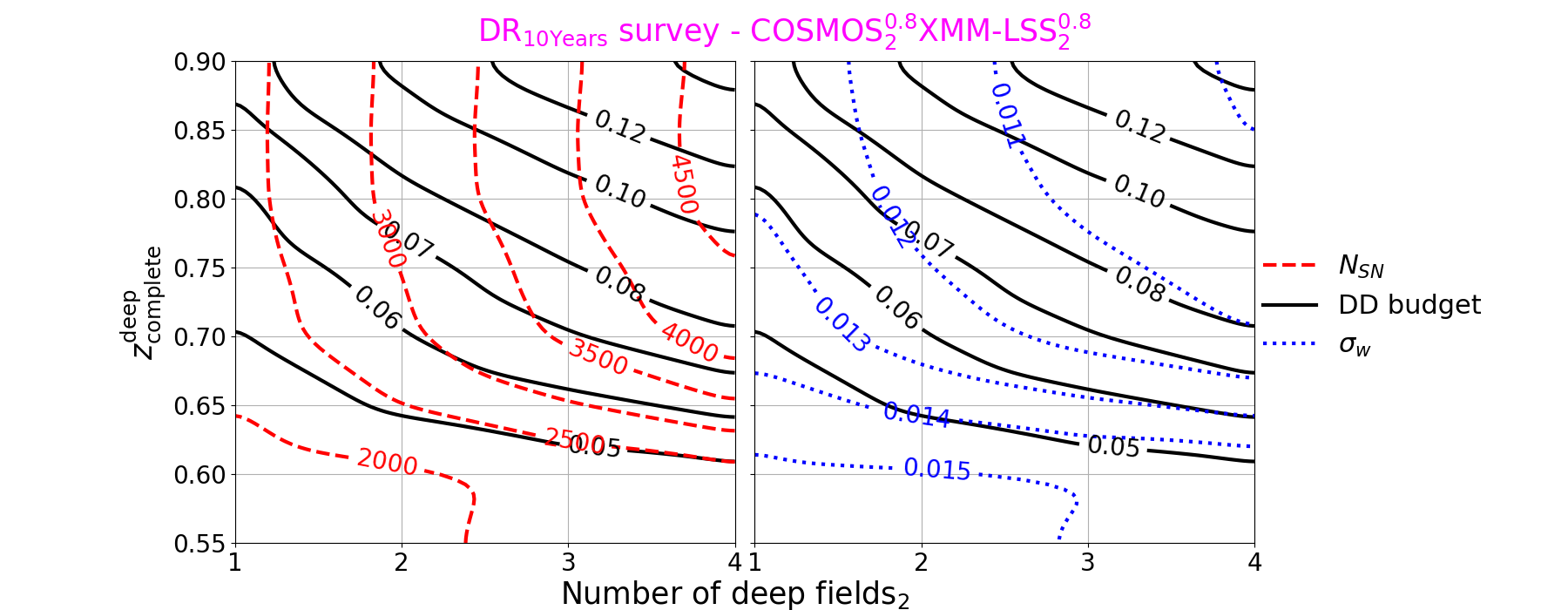}
  \end{minipage}
 \hfill
   \begin{minipage}[b]{0.9\textwidth}
    \includegraphics[width=\textwidth]{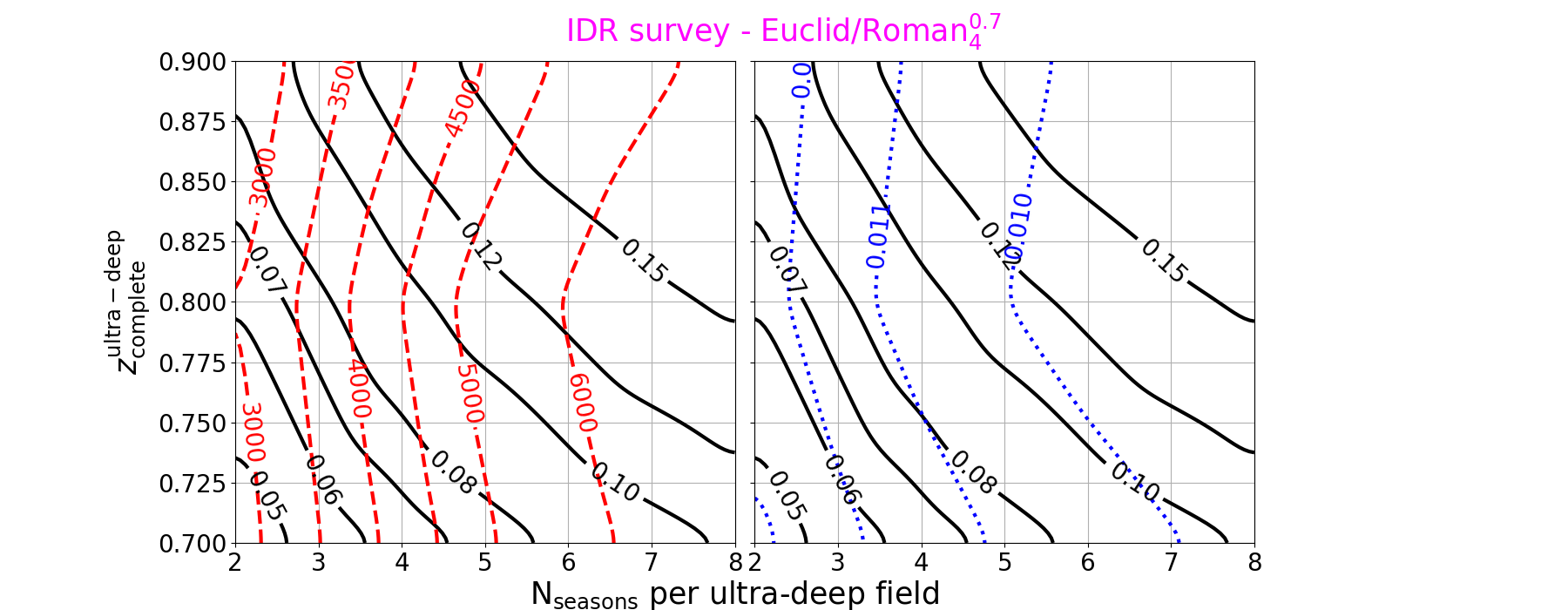}
  \end{minipage}
 \hfill
  \caption{\nsntot~(red dashed lines), DD budget (black solid lines) and \sigdew~(blue dotted lines, left plot) for three types of surveys: Deep Universal (top), Deep Rolling 10 years (middle) and Intensive Deep Rolling (bottom). Subscripts correspond to the number of seasons of observation (6 months season length) and superscripts to the redshift completeness. The number of $y$-band visits is less than 20 (per observing night). Only statistical uncertainties are included.}\label{fig:cosmo_contour}
  
\end{figure*}

In the Deep Universal (DU) survey, all the DDFs are observed for ten years with a similar cadence, season length and \zcomp~(i.e. the same number of visits per band and per observing night). The budget is the major factor limiting the redshift completeness of the survey (Fig. \ref{fig:cosmo_contour}, top). A budget higher than 15$\%$ is required to reach \zcomp~$\lesssim$~0.65 with 5 DDFs (and 2 pointings for \adfs). This configuration leads to a high number of well-sampled \sne~($\sim$ 14,000) and to the best cosmological constraints (\sigdew~$\sim$~1\per). \zcomp~and \nsntot~dramatically decrease for a budget lower than 5\per~and the cosmological measurements get significantly worse (\sigdew~$\geq$~2\per).

\paragraph{Cadence of observation}
Reaching \zcomp$\sim$0.65 can be achieved with 41 visits per night of observation with a 1-day cadence (\autoref{sec:opti}). Up to 4 fields (because of its northernmost location, \cosmos~is visible independently of the other DDFs) may have to be scanned for some nights, which corresponds to 1 hour and 42 minutes of observation. Moving to a 3-day cadence would require 123 visits per field and per night of observation. In that case the scanning of 4 fields (5 pointings) has to be spread over 3 nights with at least two fields observed per night. This corresponds to about 2 hours of observation.

\subsection{Deep Rolling Surveys}\label{sec:deep_rolling_survey}

Observing 5 fields for ten years up to \zcomp~$\sim$~0.9 would certainly give access to a large sample of well-measured \sne~(around 19k) but also to an unrealistic scenario (N$_{visits}^{DD}~\gtrsim~\mathrm{N}_{visits}^{WFD}$). The only way to reach higher \zcomp~while remaining within a reasonable budgetary envelope is to reduce the number of fields to be observed and/or the number of seasons of observation. We propose the Deep Rolling (DR) strategy characterized by a limited number of seasons of observation per field (at least 2) and a large number of visits per observing night (more than 130 for higher \zcomp).

A realistic large scale high-$z$ DR survey is characterized by: (a) a high cadence of observation (one day), (b) a rolling strategy (with a minimal of two seasons of observation per field), and (c) two sets of fields, ultra-deep (\cosmos, \xmm, with \zcomp~$\gtrsim$~0.8) and deep (\adfs, \cdfs, \elais, with \zcomp~$\gtrsim$~0.7) fields. We study two DR scenarios in the following, characterized by a minimal number of seasons of observation per field or by a minimal number of fields to be observed. 

\subsubsection{Deep Rolling 10 Years}
In this scenario all the DDFs are observed for two seasons. The results of the triplet (budget, \sigdew, \nsntot) as a function of the number of DDFs and redshift completeness are given on Fig. \ref{fig:cosmo_contour} (middle). It seems difficult, with a budget lower than 5$\%$, to perform cosmological measurements of \dew~with an accuracy better than 1.5$\%$. A sample of $\sim$~4000-4500 well-measured \sne~with a reasonable budget of $\sim$~8$\%$ would lead to \sigdew~$\sim$~1.1$\%$. 

This scenario appears to have two essential weaknesses. The number of seasons of observation per field is low. Periods of bad weather could affect the progress of the survey and the quality of the \sne~sample. What is more, this survey requires a precise timeline that may be difficult to tune (see an example below).

\begin{figure}[htbp]
\begin{center}
  \includegraphics[width=0.5\textwidth]{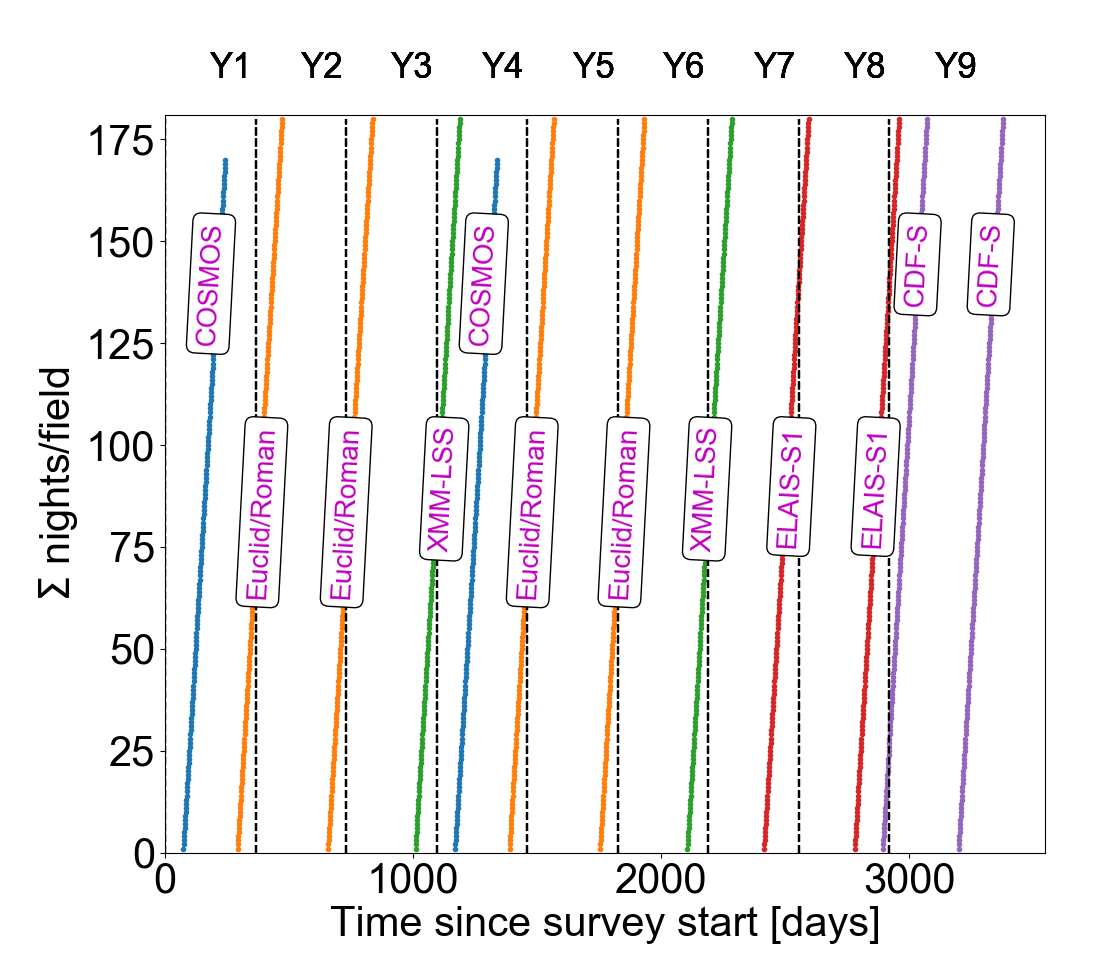}
  \caption{ Cumulative sum of the number of nights (per field and per season) as a function of the time since survey start (assumed to be late 2023).  The following sequence of observations is considered: \cosmos, \adfs~(x2), \xmm, \cosmos, \adfs~(x2), \xmm, \elais~(x2), \cdfs~(x2) ,  with  a  maximum  season  length  of  180  days for the deep fields,  a  cadence  of  one day, and ensuring only one field is observed per night.  The fields are required to be observable (airmass~$\leq$~1.5 and 20\textdegree~$\leq$ altitude~$\leq$~86.5\textdegree) for at least 1 hour and 5 minutes (131 visits) for the ultra-deep fields and 34 minutes (68 visits) for the deep fields.  The overlap, defined as the fraction of nights with more than one field observed during a night, is $\sim~4\%$}\label{fig:timelysequence}
\end{center}
\end{figure}

The sequence of observations (field/night) of the DR survey must fulfill a couple of constraints. LSST observation of \adfs~has to be contemporaneous with \euclid~(years 2 and 3) and with \romanspace~(years 5 and 6). Observing multiple fields per night is not optimal if the number of visits per field is high. It may jeopardize the uniformity of the WFD survey (if the total number of DD visits is too high) and have a negative impact on the regularity of the DD cadence (if a choice has to be made among the DDFs). Overlap of field observations should thus be minimized. This means that the DR survey should be deterministic with a timely sequence defined in advance. An example of the progress of a DR survey is given in Fig. \ref{fig:timelysequence} for a configuration of 5 fields and a survey complete up to $z\lesssim$ 0.8 and $z\lesssim$ 0.7 for ultra-deep and deep fields, respectively.

\subsubsection{Intensive Deep Rolling}
In this scenario a minimal number of fields are observed and the goal is to maximize the redshift completeness of the survey\footnote{It is defined as the median redshift completeness of the observed fields.}. The choice of the fields may be motivated by the following considerations: the need to explore high redshift completeness domains, and the synergy with surveys contemporaneous with Rubin Observatory operations. Fulfilling these requirements leads to the choice of three fields: \cosmos, \xmm~and \adfs. In this scenario \adfs~is observed for 4 seasons up to \zcomp $\sim$~0.7 and \cosmos~and \xmm~are observed every year with a redshift completeness of at least 0.7. Reaching higher \zcomp~domains requires to increase the number of visits per observing night ($\sim$~130 visits to reach \zcomp~$\sim$~0.8 for a 1-day cadence). The DD budget may then be used up in few years and this is why this scenario may be dubbed as ``intensive". The results of the triplet (budget, \sigdew, \nsntot) as a function of the number of seasons of observation and redshift completeness of the two ultra-deep fields are given on Fig. \ref{fig:cosmo_contour} (bottom). About 3500-4000 well-sampled \sne~are collected after 3 to 4 years of observation of \cosmos~and \xmm~up to \zcomp~$\lesssim$~0.8. The corresponding budget is 8\per~and \sigdew$\sim$1.1-1.2\per.

\subsection{Conclusion: DR~surveys yield more accurate cosmological measurements}

\begin{figure}[htbp]
\begin{center}
  \includegraphics[width=0.4\textwidth]{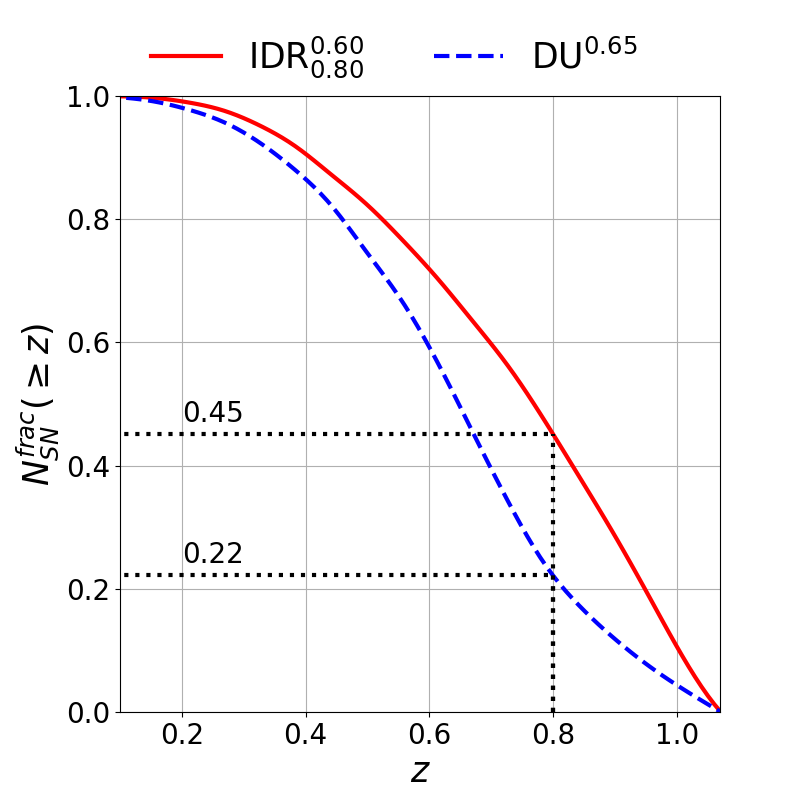}
  \caption{Fraction of \sne~($\geq~z$) as a function of the redshift ($z$) for two scenarios: \edrs{0.80}{0.60} (full red line) and \dus{0.65} (dotted blue line). The fraction of \sne~with $z~\geq~$0.8 is of 45\per~and 22\per~ for \edrs{0.80}{0.60} and \dus{0.65} surveys, respectively.}\label{fig:nsnfrac_z}
\end{center}
\end{figure}

Few conclusions can be drawn from a comparison of the surveys presented in \autoref{sec:deep_universal_survey} and \ref{sec:deep_rolling_survey}:
\begin{itemize}
    \item \textit{redshift completeness}: it is impossible to reach \zcomp $\geq$ 0.6-0.65 for \dus~surveys. This is to be explained by the DD budget envelope leading to a limited number of visits per observing night if a large number of fields are observed all seasons for ten years. This results in a low number of visits in the redder bands ($z$ and $y$) imposing a limit on \zcomp.
    \item \textit{accuracy of cosmological measurements}: under the assumption of an identical budget, \edr~surveys lead to more accurate cosmological measurements. With a DD budget of 5\per, the \dew~parameter can be measured with \sigdew~$\sim$~1.3-1.4\per~for \edr~scenarios and to \sigdew~$\geq$~2\per~for \dus~surveys. This result is mainly due to the fact that the distribution of the number of \sne~\nz~depends on the redshift completeness value of the survey (see Fig. \ref{fig:sigma_mu_nsn_bias} in \autoref{appendix:realsimu}). \edr~surveys present a higher fraction of \sne~at higher redshift compared to \dus~surveys (Fig. \ref{fig:nsnfrac_z}) and lead to more accurate cosmological measurements. 
    \item \textit{DD Budget impact}: moving from a budget of 5\per~to 8\per~would lead to a relative decrease of \sigdew~of 20-25\per, depending on the scenarios. Measuring \dew~with a precision on $~\sim~$1.1\per~requires a minimal budget of 8\per~and 15\per~for DR and DU scenarios, respectively.
\end{itemize}

\section{Realistic surveys - impact of host-galaxy redshifts}
\label{sec:realistic_surveys}
The goal of this section is draw a comparison of the performance of realistic surveys using the triplet  (budget, \sigdew, \nsntot). 
Effects of two critical aspects have to be included to get a more accurate comparison of the proposed surveys: the Malmquist bias correction and the \sne~host galaxy redshift estimation. 

Systematic uncertainties related to observational selection effects will probably account for a major part of the error budget in the era of LSST. We have considered two components related to the selection bias : a statistical contribution, related to the limited number of \sne~per redshift bin ; a bias contribution, due to the (limited) knowledge of the (\snstrech,\col) distribution of the selected \sne. We have used the G10 intrinsic scatter model (\citealt{Scolnic_2016}) where (\snstrech,\col) distributions are described by asymmetric gaussian distributions with three parameters and their uncertainties $\sigma$. We have performed simulations by individually varying each parameter of $\pm$1$\sigma$. The differences of distance modulus values w.r.t. the nominal configuration were added quadratically to provide systematic uncertainties.

Measuring cosmological parameters with a high degree of accuracy requires to minimize uncertainties of the two components of the \sne~Hubble diagram: the distance modulus and the redshift. Collecting a large sample of well-measured \sne~leading to accurate distance measurements is a guiding thread of this paper. It is achievable by optimizing the cadence of the survey, by adapting the number of visits per observing night, and by imposing selection criteria on photometric light curves. Redshifts can be derived either from the host galaxy (spectrum and/or photometric measurements) or from the spectrum of the supernova itself. Only host galaxy spectroscopic redshifts are considered in this study. We assume that \tides~will provide $\sim$~2500\footnote{This number corresponds to $\sim$~5\per of the host galaxy redshifts measured by \tides. The actual number is not known yet.} host galaxy redshifts (after 5 years) for the DD fields (\cdfs, \elais, \adfs) and that equatorial fields will benefit from \pfs~measurements ($\sim$~20,000 spectra after ten years). The current \pfs~strategy is to cover 5 \degsq~(4 PFS FoV) for equatorial fields to accumulate $\sim~$12,000 host galaxy spectroscopic redshifts over 10 years. The remaining DD area would be observed once the LSST survey is completed. The fraction of \sne~expected to have secure redshift measurements is taken from Fig. 1 of \citealt{mandelbaum2019widefield}.

\begin{figure*}[!tbp]
\begin{center}
    \includegraphics[width=0.70\textwidth]{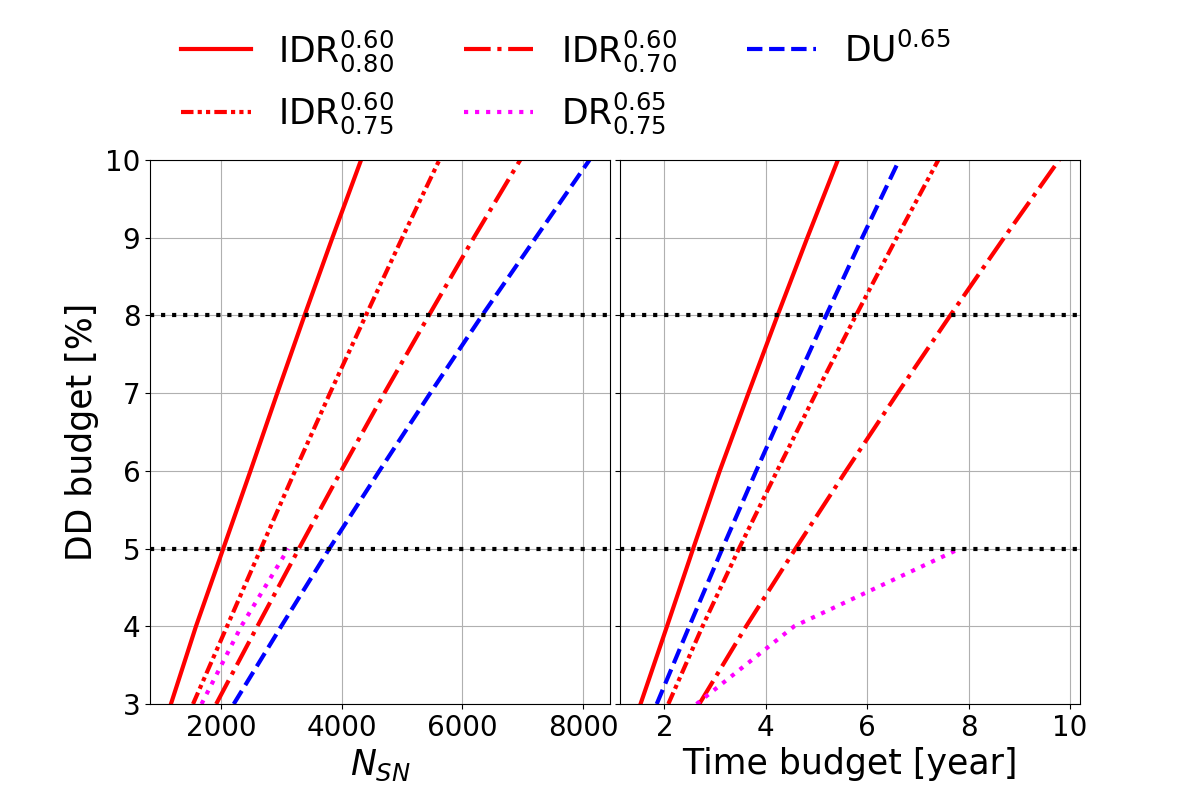}
  \caption{DD budget as a function of the total number of well-sampled \sne~(left) and time budget (right) for a set of Intensive Deep Rolling (IDR), Deep Rolling (DR) and Deep Universal (DU) surveys. Subscripts (superscripts) correspond to the redshift completeness of ultra-deep (deep) fields. Black dotted lines correspond to DD budgets of 5\per~and8\per.}\label{fig:survey_compare_a}
  \end{center}
\end{figure*}


We have chosen a large set of surveys among the possible configurations presented above (observing strategy parameters are listed in Tab. \ref{tab:surveyparam}):
\begin{itemize}
    \item Deep Universal survey (\duex): 5 DD fields - \cosmos, \xmm, \cdfs, \elais, \adfs~ - are observed every season with \zcomp~$\in~$[0.60,0.80] ($\Delta z~=~0.05)$
    
    \item Deep Rolling 10 years (\drall): 5 DD fields -~\cosmos, \xmm, \cdfs, \elais, \adfs~- are observed for two seasons each according to the timeline defined in Fig. \ref{fig:timelysequence}. The redshift completeness ranges are [0.60,0.80] and [0.50,0.75] ($\Delta z~=~0.05)$ for ultra-deep and deep fields, respectively.
    
    \item Intensive Deep Rolling (\edr): 3 DD fields are considered: two ultra-deep fields (\cosmos,\xmm) with \zcomp~$\in~$[0.70,0.75,0.80] and one deep field (\adfs) with \zcomp~$\in~$[0.50,0.70] ($\Delta z~=~0.05)$ 
    . Ultra-deep fields are observed every year and \adfs~during four seasons.
\end{itemize}

\begin{figure}[!tbp]
\begin{center}
    \includegraphics[width=0.52\textwidth]{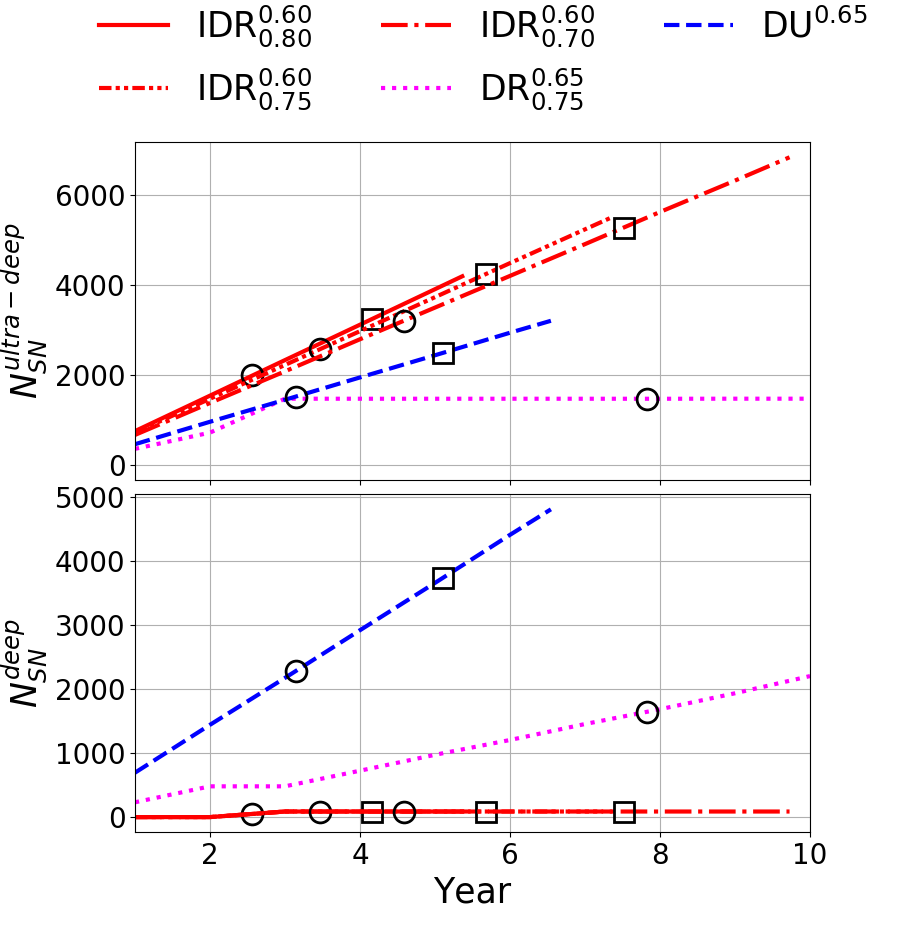}
  \caption{Number of well-measured \sne~observed in the DD survey (top: \sne~observed in ultra-deep fields, bottom: \sne~observed in deep fields) as a function of time (year of survey) for deep fields (\cdfs, \elais, \adfs) (left) and ultra-deep fields (\cosmos, \xmm) (right). The black circles (squares) correspond to a budget of 5\per~(8\per).}\label{fig:nsn_time}
\end{center}
\end{figure}

The budget as a function of the total number of well-sampled \sne~and the time budget are presented in Fig. \ref{fig:survey_compare_a}. As expected the larger number of \sne~is provided by DU surveys and the minimal number by IDR scenarios with a ratio $N_{SN}^{DU}/N_{SN}^{IDR_{0.80}} \simeq$ 1.8. The lowest time budget is obtained with an IDR scenario with a high \zcomp~(0.8) for ultra-deep fields: the 5\per~budget limit is reached after $\sim$~2.6 years.

The number of well-measured \sne~observed in the DD survey is given on Fig. \ref{fig:nsn_time} for deep (\cdfs, \elais, \adfs) and ultra-deep (\cosmos, \xmm) fields. IDR scenarios lead to samples mostly composed of \sne~observed in equatorial fields (the total number of \sne~provided by non-equatorial fields is of $\sim~$100). Increasing the DD budget from 5\per~to 8\per~lead to an increase of the the \sne~sample by a factor of $\sim~$1.6.

\begin{table*}[htbp]
  \caption{Observing strategy parameters for a representative set of optimized surveys. The cadence of observation is of one day and the season length of 180 days (max).}\label{tab:surveyparam}
  \begin{center}
    \begin{tabular}{c|c|c|c|c|c}
      \hline
      \hline
      \multicolumn{2}{c|}{Observing Strategy} & & & & \\
      type & name & Fields & \zcomp & N$_{seasons}$/field & \nvisits \\ 
           &      & & &  & $g/r/i/z/y$ \\ 
     \hline 
     \multirow{2}{*}{Deep Universal} & \multirow{2}{*}{\dus{0.65}} & \cosmos, \xmm & \multirow{2}{*}{0.65} & \multirow{2}{*}{10} & \multirow{2}{*}{2/9/10/15/3} \\
         &  & \elais, \cdfs, \adfs & & & \\
      \hline
      \multirow{2}{*}{Deep Rolling}  & \multirow{2}{*}{\drs{0.75}{0.65}} & \cosmos, \xmm & 0.75 & \multirow{2}{*}{2}  &  2/9/26/35/16\\
               &  & \elais, \cdfs, \adfs & 0.65 & & 2/9/10/15/3\\
      \hline
         \multirow{6}{*}{Intensive Deep Rolling}& \multirow{2}{*}{\edrs{0.70}{0.60}} & \cosmos, \xmm & 0.70 & $\geq$2 &  2/9/20/29/7 \\
               &   & \adfs & 0.60 & 4 &  2/9/1/1/\\\cline{2-6}
       & \multirow{2}{*}{\edrs{0.75}{0.60}} & \cosmos, \xmm & 0.75 &  $\geq$2 &  2/9/26/35/16\\
               &  & \adfs & 0.60 & 4 & 2/9/1/1/\\\cline{2-6}
          & \multirow{2}{*}{\edrs{0.80}{0.60}} & \cosmos, \xmm & 0.80 & $\geq$2  &  2/9/37/52/21\\
               &  & \adfs & 0.60 & 4 & 2/9/1/1/\\
      \hline
      \end{tabular}
  \end{center}
\end{table*}

\begin{figure*}[!tbp]
  \begin{minipage}[b]{0.52\textwidth}
    \includegraphics[width=\textwidth]{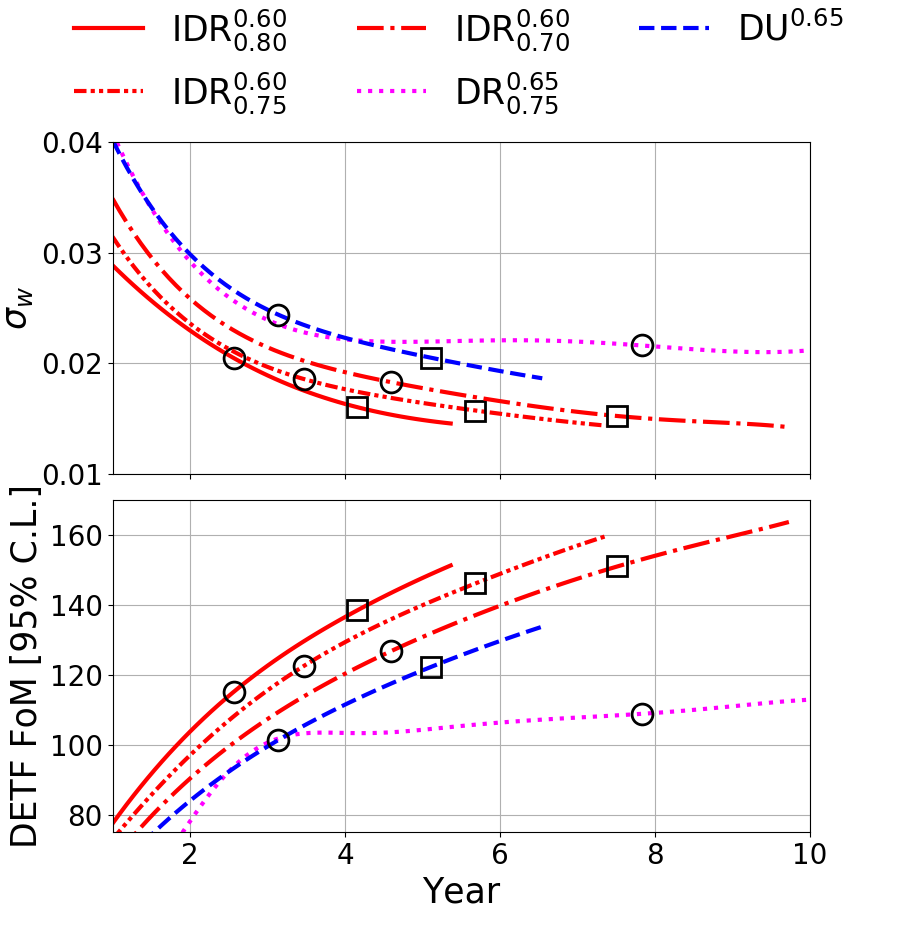}
  \end{minipage}
 \hfill
  \begin{minipage}[b]{0.52\textwidth}
    \includegraphics[width=\textwidth]{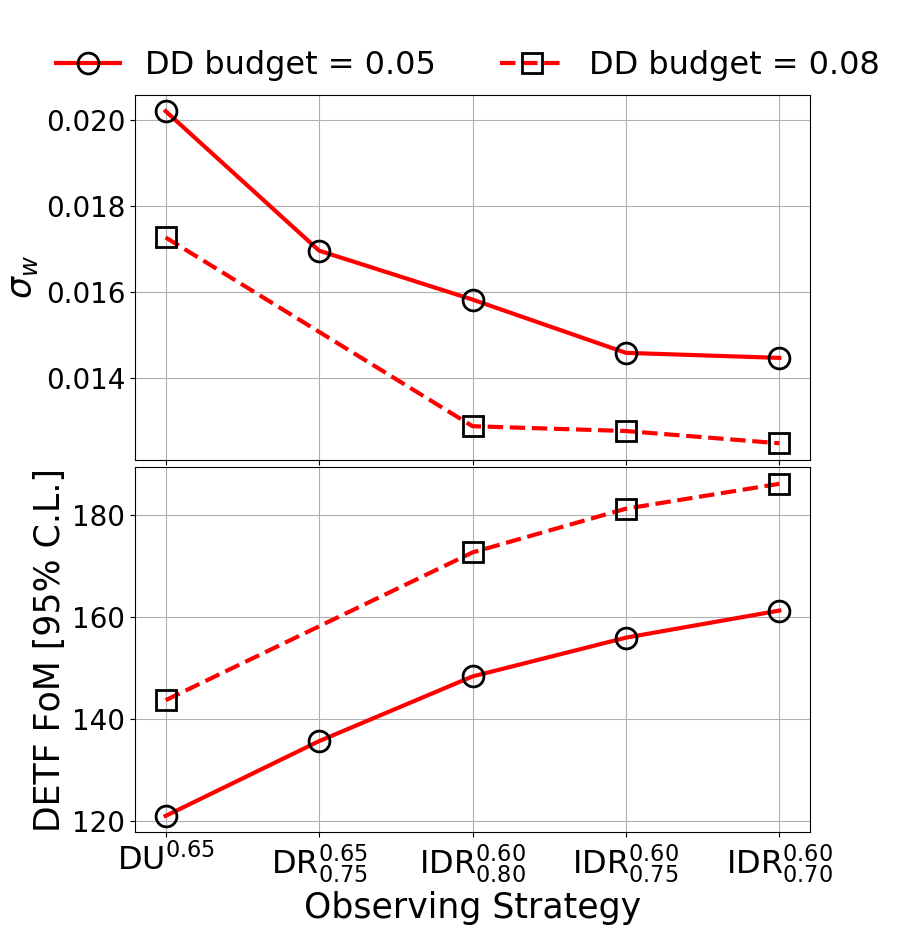}
  \end{minipage}
  \caption{Left: \sigdew~(top) and DETF FoM (bottom) a function of time (year of survey) for scenarios with a DD budget lower than 10\per. The black circles (squares) correspond to a budget of 5\per~(8\per). Right: \sigdew~(top) and DETF FoM (right) after ten years for DD strategies considered in this paper and and for DD budgets of 5\per~(solid line) and 8\per~(dashed line).}\label{fig:sigmaw_detfom}
\end{figure*}

Since only \sne~with host spectroscopic redshifts are considered, the cosmological metric \sigdew~values depend on the spectroscopic scenario (i.e. the number of host spectroscopic measured per year for \pfs~and \tides) and on the budget. The variation of \sigdew~as a function of time (year of survey) is given in Fig. \ref{fig:sigmaw_detfom} (top left) assuming the current \pfs~strategy described above. IDR scenarios tend to lead to more accurate cosmological metric estimation for the same budget. Taking into account the whole set of spectroscopic redshifts provided by \pfs~($\sim~$20,000) lead to the results of Fig. \ref{fig:sigmaw_detfom} (top right). Significant differences are observed between IDR and DU scenarios and relative decrease of \sigdew~of about 30\per~for a budget of 5\per~can be observed.


Measuring cosmological parameters with a high degree of accuracy with supernovae requires to observe a large sample of well-measured \sne~in the full redshift range [0.01,1.1]. The results of Figs \ref{fig:sigmaw_detfom} (top) indicate that the shape of the \nsntot($z$) distribution is critical to achieve low \sigdew~values. The profile of \nsntot($z$) is affected by the Malmquist bias for $z~\geq~$\zcomp~and the fraction of well-measured \sne~at high redshift depends on the detection threshold \zcomp~(Fig. \ref{fig:nsnfrac_z}). 
Between two surveys collecting the same number of well-measured \sne,  the most accurate cosmological measurements are achieved by surveys characterized by the highest \zcomp~(Fig. \ref{fig:sigmaw_detfom}, top right). Low \zcomp~surveys have to collect a higher number of well-measured \sne~to achieve the same accuracy.  

For each of the scenarios considered in this section we have estimated the figure of merit defined by the Dask Energy Task Force \citep{albrecht2006report}:
\begin{equation}
    \mathrm{DETF~FoM} = \frac{\pi}{A}
\end{equation}
where $A$ is the area of the confidence ellipse defined by:
\begin{equation}
    A = \pi\Delta \chi^2 \sigma_{w_0} \sigma_{w_a}\sqrt{1-\rho^2}
\end{equation}
with $\Delta \chi^2$~=~6.17~(95.4\per~C.L.). $\rho$ is the correlation factor equal to Cov($w_0,w_a$)/($\sigma_{w_0} \sigma_{w_a})$. $w_0$ and $w_a$ are the parameters of the Chevallier-Polarski-Linder (CPL) model of the dark energy equation of state \citep{CP2001,Linder2003}:
\begin{equation} \label{eq:cpl}
w = w_0+w_a{{z}\over{1+z}}
\end{equation}
The cosmological parameters (\omgam,$w_0$,$w_a$) have been estimated from the minimization of Eq. \ref{eq:chi_square} (modified to account for the definition of $w$ in Eq. \ref{eq:cpl}) and a prior has been added on the \omgam~ parameter (with $\sigma_{\Omega_m}~=~$0.0073~\citealt{Planck2020}). 

The $2-\sigma$ figure of merit of the scenarios considered in this section are given on Fig. \ref{fig:sigmaw_detfom} (bottom left and right). The conclusion is the same as above: IDR scenarios lead to the highest FoM in comparison with DU surveys (+25\per~between IDR$_{0.80}$ and DU$^{0.65}$ for a DD budget of 5\per). This result reflects the above-mentioned dependence of the shape of \nsntot($z$) on the redshift completeness of the survey.

One of the main conclusions of the studies presented in this section is that the design of an optimal LSST DD mini-survey for cosmology with \sne~has to include external critical datasets such as precise host redshift measurements to be realistic. Further studies are needed to assess the impact of spectroscopic datasets from \pfs, \tides, and of synergy with \euclid~and \romanspace, on the scenarios proposed in this paper. Optimizing the use of limited spectroscopic resources requires to select DD surveys that deliver a sample containing a low number of well-measured \sne~while leading to accurate cosmological measurements. IDR (high \zcomp) surveys fulfill both criteria.

The large number of surveys presented in this section were achieved with optimal observing conditions (regular cadence, median \fivesig~depth). Additional simulations are required to assess the effect of realistic observing conditions such as variations of \fivesig~depth values or of the Moon brightness. The results of such studies might require to tune the parameters of the surveys proposed in this paper. 

\section{Gap/budget recovery}\label{sec:gaprecovery}
\begin{figure}[htbp]
\begin{center}
  \includegraphics[width=0.5\textwidth]{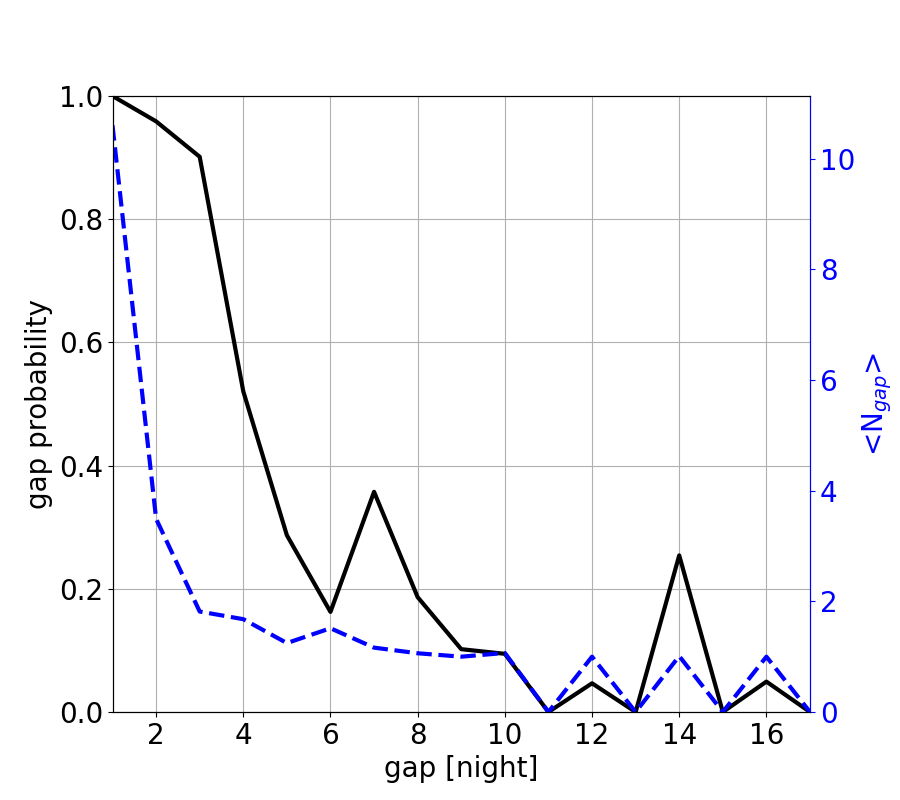}
  \caption{Probability to have gaps (full line, left $y$-axis) and mean number of gaps (dotted line, right axis) as a function of the night gap estimated with a sliding window of 180 days using the LSST baseline simulation dubbed baseline\_nexp1\_v1.7\_10yrs.}\label{fig:gapbaseline}
\end{center}
\end{figure}

All the results presented above assume a regular cadence with median observing conditions (i.e. median $m_5$) and no translational dithering. But it is known that the LSST survey will be affected by gaps originating from telescope downtimes related to maintenance periods and to poor observing conditions (clouds). The probability to have low gaps (few nights) is high (more than 80$\%$ for gaps lower than 3 nights) and mainly due to dome closed periods (Fig. \ref{fig:gapbaseline}). Larger gaps are explained by telescope maintenance times and are not exceptional: the probability to have a gap of 14 nights is of about 25 $\%$. It has been shown (\autoref{sec:analysis}) that gaps affect the redshift completeness and the total number of well-measured \sne, but also the DD budget if the season length is fixed. 

\begin{figure*}[!htbp]
\begin{center}
  \includegraphics[width=0.75\textwidth]{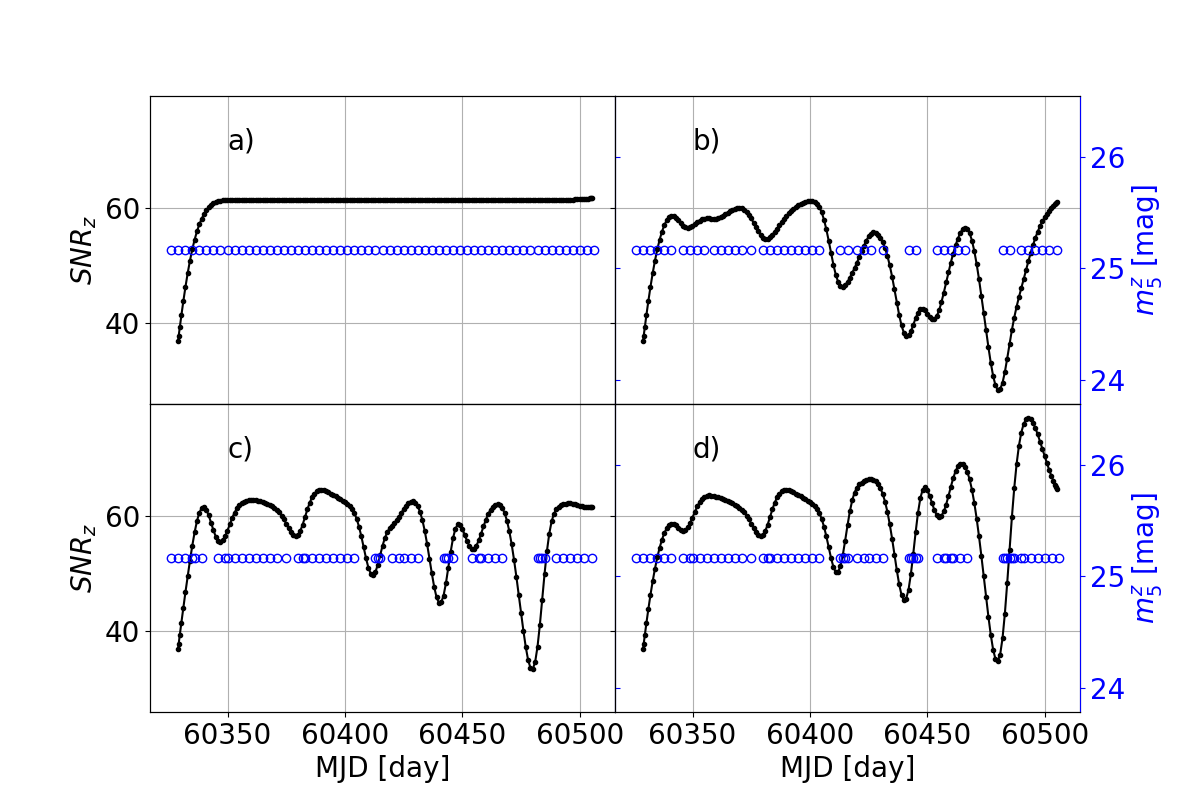}
  \caption{Gap tracker metric as a function of MJD. The full black lines correspond to \snrcom{z}~and the blue points to observations ($m_5$). Four cases are represented: 3-day cadence surveys with no gaps (a), with gaps (b), with gaps and a recovery method based on the number of visits (c), with gaps and a recovery method based on the gap tracker metric (d).}\label{fig:snrtime_recovery}
\end{center}
\end{figure*}

The metrics used up to this point were estimated using the whole set of data. The impact of the gaps is estimated at the end of the survey. It may be interesting to define a metric that would trace gap effects on a nightly basis. We propose as a gap tracker metric the $z$-band SNR (\snrcom{z}) of a medium \sne~(i.e. with (\snstrech,\col)=(0.0,0.0)) at a redshift $z\sim$0.6. This \sne~is characterized by \daymax=\mjdnight~where \mjdnight~is the Modified Julian Date of a given night. \snrcom{z}~is estimated from the rising part of the \sne~light curve. The value of \snrcom{z}~is constant for regular cadences (Fig. \ref{fig:snrtime_recovery}a) and decreases with gaps (Fig. \ref{fig:snrtime_recovery}b).

\begin{table}[!htbp]
  \caption{Variations of budget, \zcomp, \nsntot, and season length(sl) for a set of surveys with gaps and three recovery methods. The reference values are taken from a survey with a regular cadence of three days (no gaps) and a 6 months season length.}\label{tab:snrtimerecov}
  \begin{center}
    \begin{tabular}{c|c|c|c|c}
      \hline
      \hline
      Survey & $\Delta$Budget & $\Delta$\zcomp & $\Delta$\nsntot & $\Delta$sl\\
      & $\%$& & $\%$ & days\\
      \hline
      gaps & -25& -0.03& -18 & 0\\
      season length extens. &0 &-0.02 &+11 & +45\\
      nightly $\Delta N_{visits}$& 0 &-0.01 & -6& 0\\
      gap tracker thresh. & -8 & -0.02& -9& 0\\
      \hline
      \hline
      \end{tabular}
  \end{center}
\end{table}

A survey with the gap distribution of Fig. \ref{fig:snrtime_recovery}b lead to a decrease of \zcomp~of about 0.03 and to a loss of about 18\per~of well-measured \sne. The budget is also affected (-25\per). Three recovery methods can be used to get (\zcomp, \nsntot, budget) values close to a survey with a regular cadence (as in Fig. \ref{fig:snrtime_recovery}a). The first approach would consist in recovering the initial budget (i.e. the total number of observing night) by adding, at the end of each season, a number of observing nights corresponding to the number of downtime nights. This method has limited effects on \zcomp~but lead to an increase of the size of the well-measured \sne~due to the increase of the season length (Tab. \ref{tab:snrtimerecov}). The second approach relies on the comparison, at the beginning of a night, between the number of visits $N_{obs}$ and the number of expected visits $N_{exp}$ corresponding to a survey without gaps. Observations are added while $\Delta N_{visits} = N_{obs}-N_{exp}<0$ (Fig. \ref{fig:snrtime_recovery}c). This method leads to a complete recovery of the budget and to a minimal loss of well-measured \sne~(Tab. \ref{tab:snrtimerecov}). The last approach exploits the fact that the gap tracker metric strongly decreases with gap widths. The recovery is made by adding observations if \snrcom{z}~values are lower than a threshold defined from the survey without gaps (Fig. \ref{fig:snrtime_recovery}a). A partial recovery can be obtained from this method (Fig. \ref{fig:snrtime_recovery}d and Tab. \ref{tab:snrtimerecov}) which requires some tuning of the threshold value.

A closer look at the results of Tab. \ref{tab:snrtimerecov} indicates that one of the best ways to recover from gap effects is to extend the season length of observations. This method could nonetheless not be applicable for all the DDFs considered in this paper. Because of their northernmost positions, \cosmos~and \xmm~have the lowest season lengths (Fig. \ref{fig:seasonlength_nvisits_new}) and extensions may not be possible (depending on the number of visits). In that case the second ($\Delta N_{visits}$ comparison) or third method (gap tracker) should be preferred.







\section{Conclusions}
\label{sec:conclusion}
In this paper we have presented a three-phase study to assess the impact of the LSST Deep-Drilling mini-survey on the size and depth of a sample of well-measured \sne: (a) thorough analysis of DD strategy proposed by LSST, (b) development of a method to probe higher redshift completeness domains and (c) proposal of a set of optimized DD surveys.
\begin{enumerate}[label=(\alph*)]
\item A comprehensive analysis of LSST simulations has been achieved on a subset of representative DD scenarios. We have studied the impact of cadence, gaps, and translational dithering using the metric (\zcomp,~\nsntot). It was shown that reaching a redshift completeness higher than 0.55-0.65 is difficult with a reasonable ($\sim$~5\per) budget allocation.

\item Reaching higher redshift completeness requires increasing the signal-to-noise ratio of the photometric light curves, while considering band-flux distribution ($z$-dependent), cadence, and observing conditions. We have proposed a method providing the relationship between the optimized number of visits {\it per band}  and the redshift completeness. We have used this result to design a set of realistic strategies.

\item Two classes of optimized surveys have been studied. In the Deep Universal strategy all the DDFs are observed with the same cadence of observation, season length and number of visits per observing night (i.e. same \zcomp). The Deep Rolling strategy, where two classes of fields are defined (deep and ultra-deep), aims at probing higher \zcomp~domains. Host galaxy spectroscopic measurements from surveys contemporaneous with LSST have been included to design realistic optimized surveys.
\end{enumerate}

The results shown in this paper represent a first step towards the design of an optimal LSST DD survey for cosmology with \sne. Simulations of the DD survey with the LSST scheduler would help in quantifying the impact of realistic observing conditions (cadence, gaps, \fivesig~depth, Moon brightness) on the proposed surveys. Additional studies are required to fully benefit from spectroscopic resources (DESI, \pfs, \tides) and to optimize the synergy with \euclid~and \romanspace.

The studies presented in this paper lead to following main conclusions:
\begin{enumerate}
    \item \textit{Simulated LSST surveys do not lead to precision cosmology} \\
    The DD surveys proposed by LSST lead to a sample of well-measured \sne~too shallow to measure cosmological parameters with a high degree of accuracy. The redshift completeness of the survey is too low. Large inter-night gaps lead to low cadences of observation and to a dramatic decrease in the \sne~sample size (up to 30\per). Having a deterministic scheduler would provide significant improvements in the quality of the survey, in terms of cadence regularity or season length. The scheduler could include metrics (gap recovery mechanism) to monitor the DD survey on a nightly basis and to correct for gaps so as to achieve a high quality observing strategy for \sne~cosmology with LSST. Large translational dithers reduce the DDF area observed with high cadence and lead to a dramatic decrease of the number of well-measured \sne~for low cadence surveys. 
    
    \item \textit{Intensive Deep Rolling for precision cosmology}\\
    Of the variety of optimized DD scenarios studied in this paper, Intensive Deep Rolling (IDR) surveys lead to the most accurate cosmological measurements. They are characterized by a minimal configuration of 3 fields, with two ultra-deep (\cosmos~and \xmm~up to \zcomp$~\sim~$0.8) and one deep (\adfs~up to \zcomp$~\sim~$0.6) fields. These scenarios require a high cadence of observation (every night), a large number of visits per night (about 130) and could be achieved in few years. \\
    Measuring cosmological parameters with a high degree of accuracy requires to have a sample of well-measured \sne~with a significant fraction of supernovae at higher redshifts. This fraction increases with the redshift completeness of the sample. IDR are characterized by highest \zcomp~ and lead to the best figure of merit. 
    
    \item \textit{Accurate $z_{host}$ critical for precision cosmology} \\
    The accuracy of the cosmological measurements with \sne~is crucially dependent on the precision of the host galaxy redshifts ($x$-axis of the Hubble Diagram). Spectroscopic resources will be critical for higher redshift ($z~\gtrsim~$0.7) \sne~that significantly contribute to the measurement of cosmological parameters such as $(w_0,w_a)$. \pfs~is currently the only survey able to provide a significant number of spectra (few thousands) in the range $z~\in~$[0.7,1.1] during LSST era. This explains why the best strategies proposed in this paper are based on the intense observation of two equatorial fields (\cosmos~and \xmm). Only half of the well-measured \sne~sample benefit from host galaxy spectroscopic measurements with the current \pfs~strategy. 
    Final cosmological measurements within three years of survey with the full \sne~sample would involve the delivery of about 800 spectroscopic redshifts by \pfs~per year with 8 PFS FoV per equatorial field.\\
    Additional studies are required for southern fields which benefit from \tides~host galaxy spectroscopic redshift measurements. The performance of optimized strategies were obtained under the assumption that the fraction of \sne~expected to have secure measurements in the deep fields is identical for WFD and DD fields (this fraction is not known for DD fields yet). This is probably pessimistic for DD fields and this fraction will probably be higher at high-$z$ (it is of $\sim$~20\per~for $z$~=~0.7 for WFD fields). The accuracy of cosmological measurements would improve with an increase of the fraction of \sne~with secure measurements of higher redshifts.
    
    \item\textit{Photometric redshifts, number of fields and budget}\\
    Using photometric redshifts to perform accurate cosmological measurements with \sne~in a Hubble diagram is a challenging task. It requires measuring \photz~with a high degree of accuracy and controlling catastrophic outlier redshifts to minimize \photz~systematics. The current LSST minimal target  \sigz$~\sim~0.02(1+z_{phot})$ (\citealt{photoz2018}) induces an error on the distance modulus higher than 0.10 mag (full redshift range). The impact of \sne~with \photz~on the measurement of cosmological parameters is in that case marginal. A lot of efforts is made to develop techniques leading to lower \sigz~(\citealt{Schmidt_2020}). But it will be difficult to reach an accuracy similar to spectroscopic measurements (\sigz~$\sim~10^{-3}$) and to have a set of \photz~redshifts with a precision corresponding to the stringent calibration requirements of LSST (\citealt{desc_srd}).\\
    It is thus difficult, at this stage, to design a set of DD surveys optimizing cosmological measurements while collecting a large set of \sne~with photometric redshifts. It would be more reasonable to opt for surveys optimizing the use of spectroscopic resources (host galaxy redshifts). These scenarios would guarantee a minimal figure of merit that could be improved by complementary surveys with sets of \sne~with \photz. All the DD fields that Rubin guarantees to observe could be included in such strategies. For example, if 8\per~of LSST time is spent on DDFs for supernovae,
    the survey could be composed of an \edrs{0.80}{0.60}~strategy using $\sim$~5\per~of the DD budget (in $\sim$~3 years) followed by a \dus{0.65} survey with the remaining 4 fields observed two seasons each. In this scenario, a minimal  DETF FoM of 150 would be guaranteed and all the DD fields would be observed.
    
\end{enumerate}

\vspace{0.cm}
\subsection*{Acknowledgments}
The DESC acknowledges ongoing support from the Institut National de Physique Nucl\'eaire et de Physique des Particules in France; the Science \& Technology Facilities Council in the United Kingdom; and the Department of Energy, the National Science Foundation, and the LSST Corporation in the United States.  DESC uses resources of the IN2P3 Computing Center (CC-IN2P3--Lyon/Villeurbanne - France) funded by the Centre National de la Recherche Scientifique; the National Energy Research Scientific Computing Center, a DOE Office of Science User Facility supported by the Office of Science of the U.S.\ Department of Energy under Contract No.\ DE-AC02-05CH11231; STFC DiRAC HPC Facilities, funded by UK BIS National E-infrastructure capital grants; and the UK particle physics grid, supported by the GridPP Collaboration.  This work was performed in part under DOE Contract DE-AC02-76SF00515. M.~Lochner acknowledges support from the South African Radio Astronomy Observatory and the National Research Foundation (NRF) towards this research. Opinions expressed and conclusions arrived at, are those of the authors and are not necessarily to be attributed to the NRF. H.~Awan acknowledges support from Leinweber Postdoctoral Research Fellowship and DOE grant DE-SC009193.\\ This paper has undergone internal review in the LSST Dark Energy Science Collaboration.\\
This research has made use of the following Python software packages: NumPy \citep{van_der_Walt_2011}, SciPy \citep{Virtanen_2020}, Matplotlib \citep{matplotlib_2007}, Astropy \citep{astropy_2018}, SNCosmo \citep{Sncosmo_2016}, Pandas \citep{McKinney2011pandasAF}. 


\vspace{0.5cm}
Author contributions are listed below. \\
Philippe Gris: conceptualization, software, analysis, writing. \\
Nicolas Regnault: conceptualization, analysis, writing.\\
Humna Awan: participated in (and led some) discussions on DD survey optimization, especially in regards to translational (and rotational) dithering of observations. Provided detailed comments on the paper during CWR, which hopefully strengthens the paper and makes its key takeaways clearer. Contributed to the general optimization of survey sims via Builder work.\\
Isobel Hook: general paper discussion/suggestions. Acted as internal reviewer, where comments led to several improvements. e.g. clarification of the underlying metric of success used in this paper (section 7).\\ 
Saurabh Jha: builder, enabling contributions include helping to develop the emphasis on "well-measured" supernovae in section 2.1 and section 4 (per-object uncertainty less than or on the order of the SN Ia intrinsic scatter); suggested some survey strategies considered in Table 3; participated in (and led some) discussion of paper content and helped oversee paper review as Observing Strategy Working Group co-convener.\\
Michelle Lochner: general paper discussion/suggestions including y-band optimization analysis and FoM implementation; edits of the draft; co-lead of the Observing Strategy Working Group, which provided context for strategies proposed in this paper, particularly the rolling strategies; builder.\\
Bruno Sanchez: general discussions, SNWG co-convener.\\
Dan Scolnic: general paper discussion/suggestions. Acted as internal reviewer, with comments on earlier sections and framing of the metrics used. Also, co-led Observing Strategy Working Group, which provided context for strategies proposed in this paper, particularly the rolling strategies.\\
Mark Sullivan: general paper discussion/suggestions. Internal reviewer during review process, providing significant/detailed comments which led to several changes in the paper.\\ 
Peter Yoachim: observing strategy simulations. 


\newpage
\appendix

 

\section{Metric (\nsncompb, \zcompb) estimation}
\label{appendix:metric}
\begin{figure*}[!htbp]
  \begin{minipage}[b]{0.52\textwidth}
    \includegraphics[width=\textwidth]{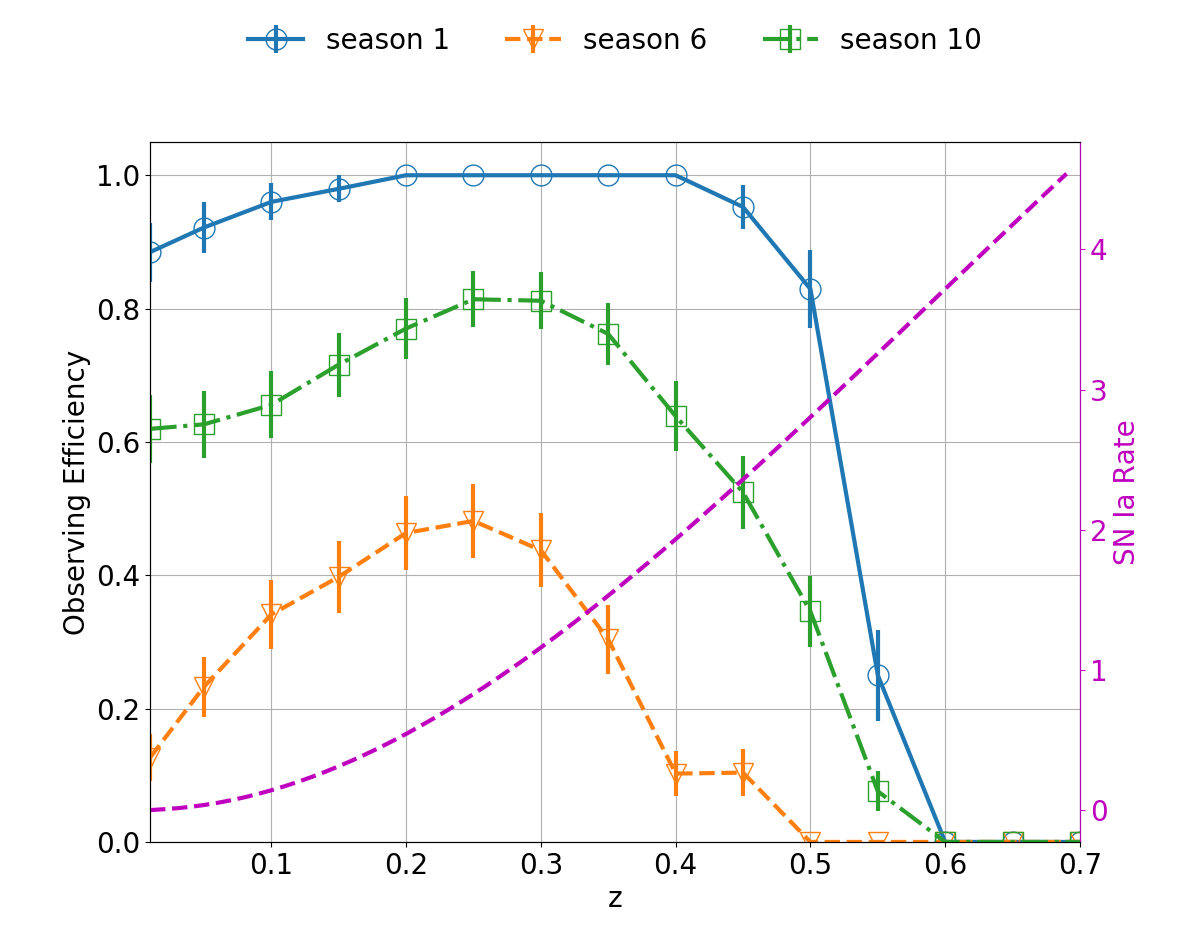}
  \end{minipage}
 \hfill
  \begin{minipage}[b]{0.52\textwidth}
    \includegraphics[width=\textwidth]{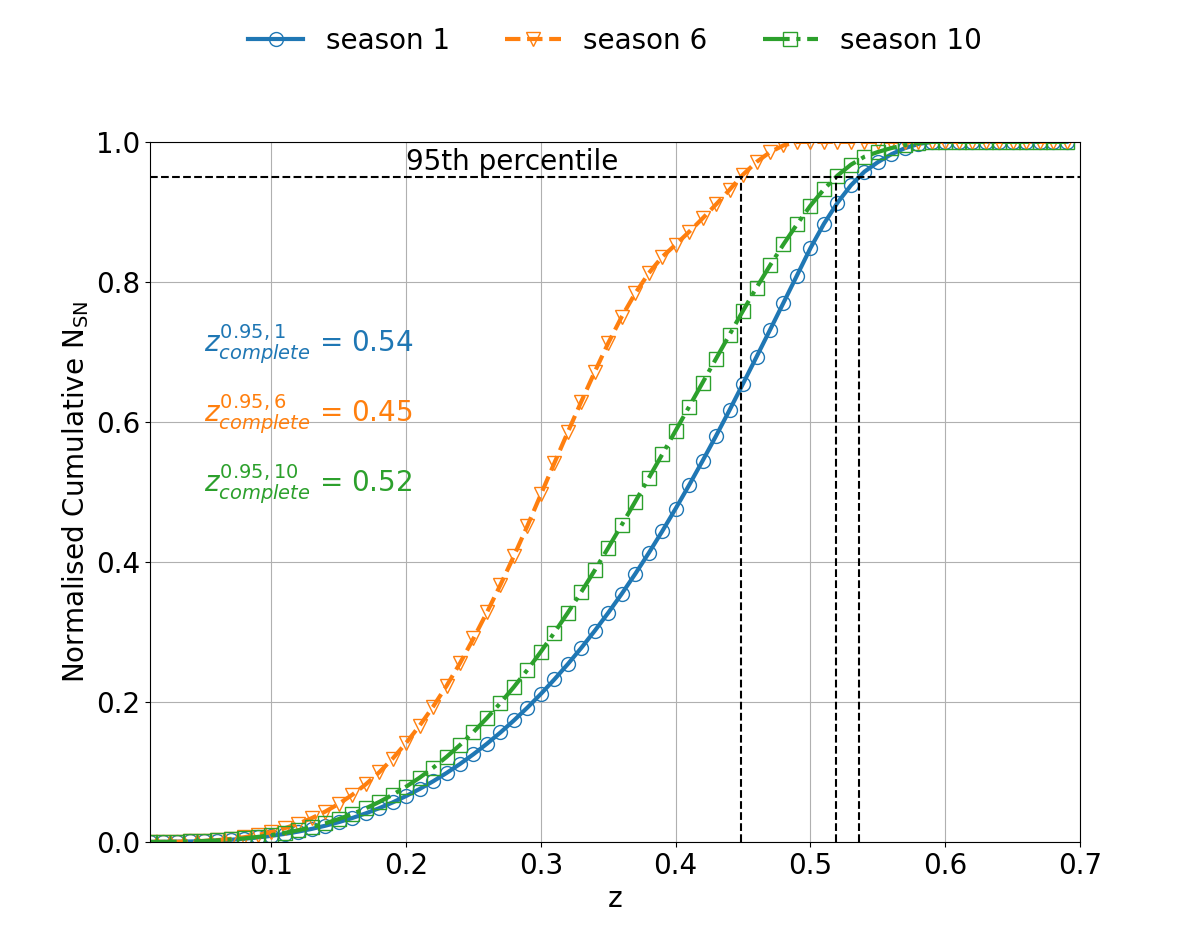}
  \end{minipage}
  \caption{Left: Observing efficiencies (color curves) and \sne~rate (black dashed curve) as a function of the redshift. Right: Normalised cumulative distribution of the number of well-measured faint supernovae as a function of redshift. The 95th percentile limit defines the \zcompb~value. These plots correspond to one HEALPix pixel of the \cdfs~field (with the number 144428) of the daily strategy.}\label{fig:method}
  
\end{figure*}

The pair metric (\nsncompb, \zcompb) is estimated from the combination of observing efficiency curves and \sne~production rate. Observing efficiencies are estimated from a set of simulated light curves of \sne. A systematic scan of the \sne~parameter space (\daymax,$z$) is performed to estimate efficiency curves with a high degree of accuracy: $z$ $\in [0.01,1.0]$ (step: 0.05) and \daymax $\in [MJD_{season}^{min}+15*(1+z),MJD_{season}^{max}-30*(1+z)]$ (step: 2 days) where MJD is the Modified Julian Date. The observing efficiency is defined as the fraction of simulated light curves fulfilling the requirements defined in Sec. \ref{sec:metrics}, per redshift bin. Example curves are given in Fig. \ref{fig:method} (left). For a regular cadence, one would expect to have the following (reference) shape: flat efficiency (close to 1) up to a completeness redshift beyond which efficiency decreases down to 0. Fig. \ref{fig:method} (left) reveals that, in practise, efficiency curves may significantly deviate from this reference shape: while results achieved in season 1 are satisfactory, it appears that observations collected in season 6 lead to poor efficiency curves (less that 50\% at max). This difference is primarily explained by the cadence of observation and the inter-night gaps that drive the sampling frequency of the light curve measurements.  

The number of well-measured \sne~ is estimated from the combination of efficiency curves and a \sne~production rate \citep{perrett}: N$_{\mathrm{SN}}(z)$ = efficiency($z$)$\times$ Rate$_{\mathrm{\sne}}(z)$. The normalised cumulative sum of N$_{\mathrm{SN}}$ is used to estimate to estimate the redshift completeness as illustrated in Fig. \ref{fig:method} (right). The results are then used as input to estimated the number of well-sampled \sne~up to $z\leq \zcompb$, \nsncompb.

\newpage
\section{Realistic simulations using $\sigma_\mu$ and $N_{SN}$}
\label{appendix:realsimu}

\begin{figure*}[!htbp]
  \begin{minipage}[b]{0.52\textwidth}
    \includegraphics[width=\textwidth]{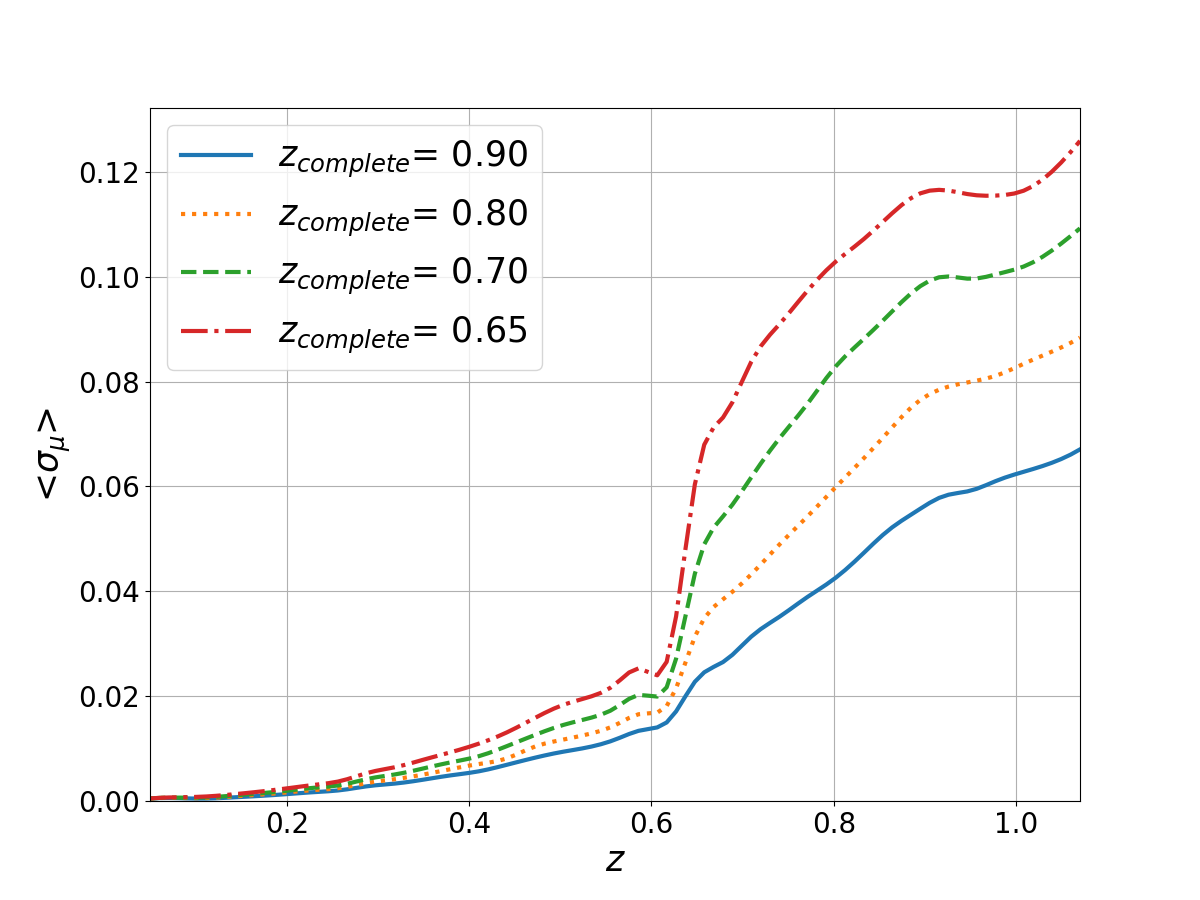}
  \end{minipage}
 \hfill
  \begin{minipage}[b]{0.52\textwidth}
    \includegraphics[width=\textwidth]{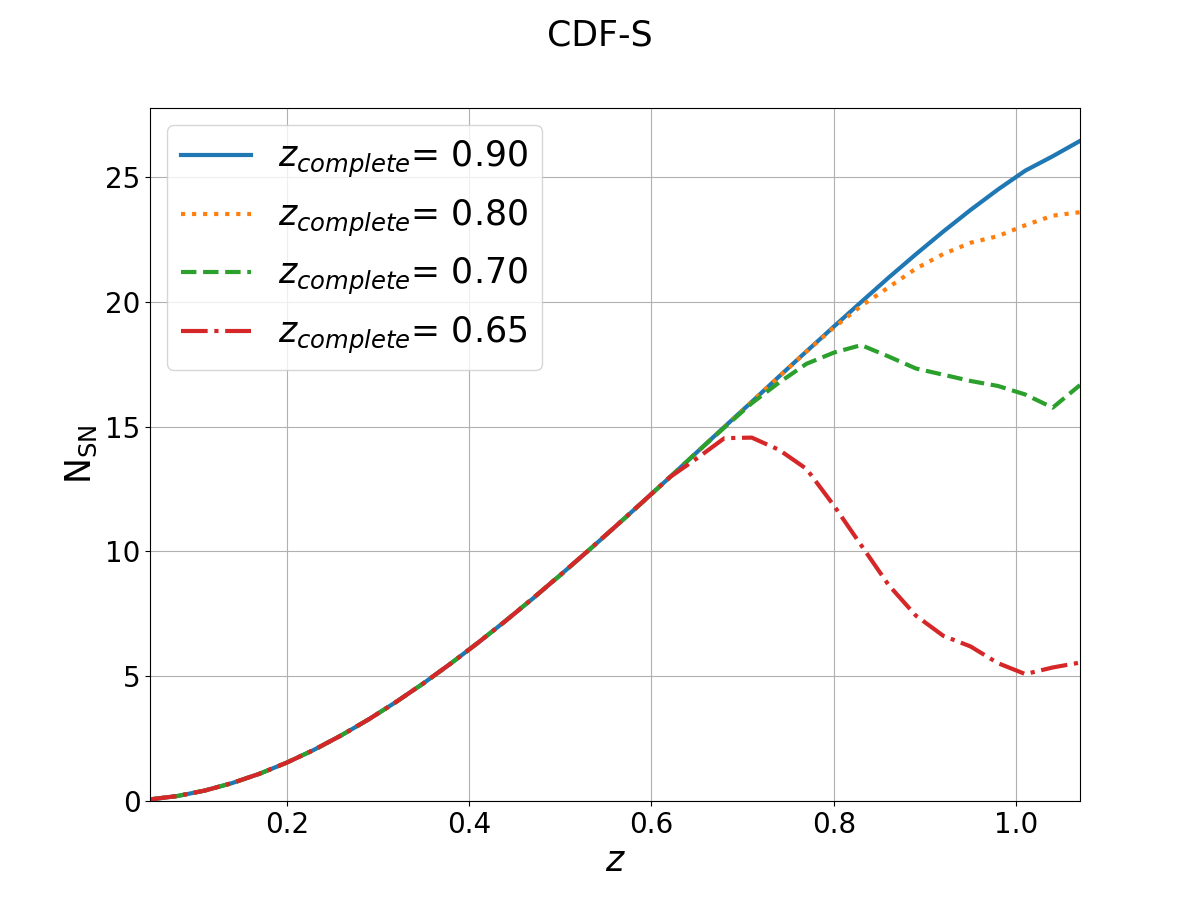}
  \end{minipage}
  \caption{Left: mean values of $\sigma_\mu$ as a function of the redshift for a set of surveys labelled by the redshift completeness. Right: number of \sne~as a function of the redshift for a set of surveys labelled by the redshift completeness  for the \cdfs~field (survey area of 9.6 \degsq~and season length of 180 days). }\label{fig:sigma_mu_nsn_bias}
  
\end{figure*}

The method used to estimate cosmological parameters (\autoref{sec:opti}) relies on the simulation of distance moduli and requires the knowledge of $\sigma_\mu$ and $N_{SN}$ as a function of $z$. These two quantities were estimated from full simulation and fit of \sne~light curves. The simulations were performed for a set of redshift completeness values, namely \zcomp$\in[0.50,0.90]$ (step: 0.05) so as to include Malmquist bias effects as illustrated by Fig. \ref{fig:sigma_mu_nsn_bias}: the dependence of $\sigma_\mu$ on \zcomp~and the decrease of $N_{SN}$~for $z~\geq~$\zcomp~is to be explained by the redshift bias.

\newpage
\bibliography{refs}

\end{document}